\documentclass[11pt,a4paper]{article}
\usepackage{amsmath,amsfonts,amssymb,amsbsy,bbold,mathtools}
\usepackage{mathrsfs,latexsym,tikz}
\usepackage{graphicx}
\usepackage{subcaption}
\usepackage{color}
\usepackage{booktabs}
\usepackage{multirow}
\usepackage{multicol}
\usepackage{alpha}
\usepackage{mathrsfs}

\usepackage{textcomp}

\usepackage[normalem]{ulem}

\begin{document}

\input{defs.sty}

\preprintno{%
\today
}

\title{\boldmath Logarithmic corrections to O($a$) and O($a^2$) effects in lattice QCD with Wilson or Ginsparg-Wilson quarks 
}


\author[uos,zppt]{Nikolai~Husung}
%
\address[uos]{Physics and Astronomy, University of Southampton,
Southampton SO17 1BJ, United Kingdom}
\address[zppt]{Deutsches Elektronen-Synchrotron DESY, Platanenallee~6, 15738 Zeuthen, Germany}

\vspace*{-1cm}

\begin{abstract}
We derive the asymptotic lattice spacing dependence $a^n[2b_0\gbar^2(1/a)]^{\hat{\Gamma}_i}$ relevant for spectral quantities of lattice QCD, when using Wilson, $\ord(a)$ improved Wilson or Ginsparg-Wilson quarks.
We give some examples for the spectra encountered for $\hat{\Gamma}_i$ including the partially quenched case, mixed actions and using two different discretisations for dynamical quarks.
This also includes maximally twisted mass QCD relying on automatic $\ord(a)$ improvement.
At $\ord(a^2)$, all cases considered have $\min_i\hat{\Gamma}_i\gtrsim -0.3$ if $\Nf\leq 4$, which ensures that the leading order lattice artifacts are not severely logarithmically enhanced in contrast to the O$(3)$ non-linear sigma model~\cite{Balog:2009np,Balog:2009yj}.
However, we find a very dense spectrum of these leading powers, which may result in major pile-ups and cancellations.
We present in detail the computational strategy employed to obtain the 1-loop anomalous dimensions already used in~\cite{Husung:2021mfl}.
\end{abstract}


\maketitle

\tableofcontents

\section{Introduction}
Today's lattice QCD simulations are reaching a point, where statistical uncertainties of several quantities obtained from the lattice become of $\ord(1\%)$ and below while smaller lattice spacings $a\simeq 0.04\,\mathrm{fm}$ become accessible.
Under these conditions predictions for precision physics are within reach, but systematic errors must be kept under control.
We focus here on the systematic error due to the continuum extrapolation.
The method of choice to describe the approach of the lattice theory to the continuum theory as $a\searrow 0$ is Symanzik Effective theory (SymEFT)~\cite{Symanzik:1979ph,Symanzik:1981hc,Symanzik:1983dc,Symanzik:1983gh}, see also~\cite[p.~39ff.]{Weisz:2010nr}.
SymEFT allows to take the quantum corrections into account, that modify the leading $a^{\nmin}$ lattice artifacts expected from classical field theory, where $\nmin$ is a positive integer and depends on the chosen lattice discretisation.
In an asymptotically free theory like QCD the leading lattice artifacts from the lattice action then have the generic form $a^{\nmin}[\gbar^2(1/a)]^{\hat\gamma_i}$, where $\gbar(1/a)$ is the renormalised coupling, $\hat\gamma_i=(\gamma_0)_i/(2b_0)$, $(\gamma_0)_i$ are the 1-loop anomalous dimensions of $(4+\nmin)$-dimensional operators and $b_0$ is the 1-loop coefficient of the $\beta$-function.
These higher dimensional operators form a minimal on-shell basis describing all lattice artifacts originating from the lattice action, which can formally be written in form of the effective Lagrangian
\begin{equation}
\Leff(x)=\L(x)+a\dlatt[1]{\L}(x)+a^2\dlatt[2]{\L}(x)+\ldots\,,
\end{equation}
where $\L$ is the Lagrangian of continuum QCD
\begin{equation}
\L=-\frac{1}{2\bare{g}^2}\tr(F_{\mu\nu}F_{\mu\nu})+\bar{\Psi}\left\{\gamma_\mu D_\mu(A) + M\right\}\Psi.
\end{equation}
$\Psi=(\psi_1,\ldots,\psi_{\Nf})^\mathrm{T}$ is a flavour vector, $D_\mu(A)=\partial_\mu+ A_\mu$ is the continuum covariant derivative with $\su(\Nc)$ algebra valued gauge field $A_\mu$, $F_{\mu\nu}=\comm{D_\mu}{D_\nu}$ is the field strength tensor and $M=\diag(m_1,\ldots,m_{\Nf})$ are the quark masses.
To obtain the leading lattice artifacts we then need to expand the SymEFT around the continuum Lagrangian such that the $\dlatt[d]\L$ are treated as operator insertions in the continuum theory, which is the common strategy in Effective Field Theories.
Here and in the following the superscript in $\dlatt[d]\L$ denotes the deviation of the canonical mass-dimension from the continuum field, e.g.
\begin{equation}
d=\big[\dlatt[d]\L\big]-\big[\L\vphantom{\dlatt[d]\L}\big].
\end{equation}
We focus on contributions from the lattice action.
This is sufficient for spectral quantities, where corrections from local fields cancel out.
For non-spectral quantities such contributions from local fields must be taken into account as well, which is beyond the scope of this paper and obviously depends on the local fields involved\footnote{For commonly used local fields a similar analysis can (and should) be performed along the lines of this paper with the sole difference that total divergence operators can no longer be discarded in the additional operator bases introduced for each local field.
See also~\cite{Luscher:1996sc} for a discussion of the minimal on-shell basis of local fields for $\nmin=1$.}\kern-0.5em.

As shown by Balog, Niedermayer and Weisz~\cite{Balog:2009np,Balog:2009yj} in the 2-d O(3) sigma model quantum corrections can spoil the approach to the continuum limit with distinctly negative values for $\hat\gamma_i$.
For this model they found $\min_i(\hat\gamma_i)=-3$, which worsens the approach from the naive $a^2$ lattice artifacts ($\nmin=2$) to something which behaves like $\ord(a)$ corrections over a long range of lattice spacings due to $\gbar^2(1/a)\sim-1/\log(a\Lambda)$, where $\Lambda$ is the intrinsic scale of the theory.
Thus computing the leading anomalous dimensions for full lattice QCD at leading order in the lattice spacing is not just a purely academic question but puts continuum extrapolations on more solid grounds by predicting the true asymptotic lattice spacing dependence.
This knowledge should then be used both in the ansatz for the continuum extrapolation and to estimate uncertainties of this extrapolation through varying the leading power in the coupling in the range predicted for the competing values~$\hat\gamma_i$.

In a previous paper~\cite{Husung:2019ytz} we discussed the special case of SU($N$) pure gauge theory with $\nmin=2$ and Wilson's lattice QCD with $\nmin=1$, which have a minimal basis of 2 or 1 operators respectively in the massless case.
In both cases the values found for $\hat\gamma_i$ are larger than zero such that discretisation errors vanish faster than the classically assumed $a^\nmin$ behaviour.
There the general concept of SymEFT theory is discussed in more detail.
It is recommended as an introduction for the reader.
In this paper we will focus on the extension to lattice QCD actions with non-perturbatively $\ord(a)$ improved Wilson quarks~\cite{Luscher:1996ug} and Ginsparg-Wilson quarks~\cite{Ginsparg:1981bj} both with $\nmin=2$, which have a significantly larger set of operators forming the minimal basis.
We also take a look at twisted mass QCD~\cite{Frezzotti:1999vv,Frezzotti:2000nk} and the connection of the minimal operator basis of untwisted QCD to the twisted case.
In particular twisted mass QCD at maximal twist ensures automatic $\ord(a)$ improvement due to an additional symmetry of the continuum Lagrangian~\cite{Aoki:2006gh,Sint:2007ug}.
While we discuss here the technicalities of the computation in some detail, we also recommend to take a look at the Letter published~\cite{Husung:2021mfl} alongside this paper, summarizing the overall results and consequences.

\section{Minimal operator basis of full QCD to \boldmath$\ord(a^2)$}
To compute the various $\hat\gamma_i$ we need the complete minimal on-shell operator basis $\op_i^{(d)}$ to express
\begin{equation}
\dlatt[d]\L=\sum_i \omega_i^\op(\bare{g}^2)\op_i^{(d)},\label{eq:matchingCoeffs}
\end{equation}
where $\omega^\op_i(\bare{g}^2)$ are the bare matching coefficients with bare continuum coupling $\bare{g}^2$.
The expansion in $\bare{g}^2$ of $\omega^\op_i$ can be determined for any chosen lattice discretisation.
Which operators must be included into the basis, depends on the symmetries realised for the lattice discretisation, i.e.~the lattice action
\begin{equation}
S_\mathrm{QCD}= S_\mathrm{G}+a^4\sum_x\bar{\Psi}(x)\hat{D}\Psi(x).
\end{equation}
Different discretisations of the lattice gauge action $S_\mathrm{G}$ have already been discussed in~\cite{Husung:2019ytz} and we rather focus here on the lattice fermion actions.
Depending on the chosen lattice Dirac operator $\hat{D}$ the symmetry constraints differ.
We restrict considerations to Wilson quarks~\cite{Wilson:1974,Wilson:1975id} and Ginsparg-Wilson quarks~\cite{Ginsparg:1981bj}.

Starting with Wilson quarks the lattice Dirac operator reads
\begin{equation}
\Dw=\frac{1}{2}\left\{\gamma_\mu(\nabla_\mu^*+\nabla_\mu)-a\nabla_\mu^*\nabla_\mu\right\}+M+a\csw(\bare{g}^2)\frac{i}{4}\sigma_{\mu\nu}\hat{F}_{\mu\nu}\,,
\end{equation}
where $\sigma_{\mu\nu}=\frac{i}{2}\comm{\gamma_\mu}{\gamma_\nu}$.
Here $\csw(\bare{g}^2)=1+\ord(\bare{g}^2)$ is the improvement coefficient for the Sheikholeslami-Wohlert (SW) term~\cite{Sheikholeslami:1985ij}.
For the definition of $\nabla_\mu$, $\nabla_\mu^*$ and $\hat{F}_{\mu\nu}$ see appendix~\ref{app:latticederivatives}.
The SW-term can remove $\ord(a)$ lattice artifacts either perturbatively~\cite{Luscher:1996sc} or non-perturbatively~\cite{Luscher:1996ug}, where the latter choice also removes double mass-dimension~5 operator insertions in the SymEFT contributing to $\ord(a^2)$.

In contrast to Wilson quarks, Ginsparg-Wilson quarks~\cite{Ginsparg:1981bj} obey in the massless limit
\begin{equation}
\acomm{\Dgw}{\gamma_5}=a\Dgw\gamma_5 \Dgw,\label{eq:GW}
\end{equation}
such that they have an exact lattice chiral symmetry~\cite{Luscher:1998pqa}.
This symmetry ensures that terms like $i/4\bar{\Psi}\sigma_{\mu\nu}F_{\mu\nu}\Psi$ are already forbidden by symmetry and the lattice artifacts of the massless theory automatically start at $\ord(a^2)$.
One particular solution to~\eq{eq:GW} are Overlap fermions~\cite{Neuberger:1997fp,Neuberger:1998wv} (we choose here the conventions from~\cite{Hernandez:1998et})
\begin{align}
\Dov(M)&=\left\{1-\frac{a}{2}M\right\}\Dov(0)+M,\\
a\Dov(0)&=\mathbb{1}-A\left(A^\dagger A\right)^{-1/2},\quad A=1-a\Dw.
\end{align}
Another solution of~\eq{eq:GW} are Domain-Wall fermions~\cite{Kaplan:1992bt,Furman:1994ky} in the limit of infinite extent of the auxiliary 5th dimension~\cite{Neuberger:1997bg}.
For finite extent of the 5th dimension, chiral-symmetry violations are only exponentially suppressed as the extent of the 5th dimension increases such that flavour symmetries of Domain Wall fermions are then reduced to the ones of conventional Wilson quarks.
In summary, the symmetry constraints for Wilson and Ginsparg-Wilson quarks are the following:
\begin{itemize}
\item Local SU($N$) gauge symmetry,
\item invariance under charge conjugation and any Euclidean reflection,
\item Hypercubic \cubicSymm{} symmetry as a remnant of broken O(4) symmetry,
\item $\SUN[\Nf]{L}\times\SUN[\Nf]{R}\times\UN[1]{V}$ flavour symmetry for massless lattice fermion actions preserving lattice chiral symmetry,
\item $\UN[\Nf]{V}$ flavour symmetry for massless (or mass-degenerate) Wilson quarks.
\end{itemize}
Due to being interested in the minimal on-shell basis we may further make use of the continuum equations of motion~(EOM)
\begin{equation}
[D_\mu(A), F_{\mu\nu}]=T^a\bare{g}^2\bar\Psi\gamma_\nu T^a\Psi,\quad \gamma_\mu D_\mu(A) \Psi=-M\Psi,\quad \bar\Psi \cev D_\mu(A) \gamma_\mu=\bar\Psi M,\label{eq:EOMs}
\end{equation}
to eliminate redundant operators~\cite{Luscher:1996sc}.
Here $T^a$ denotes the generators of the $\su(\Nc)$ colour algebra.

\subsection{Massless operator basis}\label{sec:masslessBasis}
For massless Wilson quarks with at most perturbative $\ord(a)$ improvement, one operator is required at mass-dimension~5 describing the occurring $\ord(a)$ lattice artifacts.
We previously discussed this operator~\cite{Husung:2019ytz} and list it here for completeness
\begin{subequations}\label{eq:minBasis}
\begin{equation}
\opFive_\mathrm{1}=\frac{i}{4}\bar{\Psi}\sigma_{\mu\nu}F_{\mu\nu}\Psi.
\end{equation}
For mass-dimension~6 we restate here the minimal basis of pure gauge theory~\cite{Weisz:1982zw,Luscher:1984xn}, which is of course a subset of the operator basis of full QCD,
\begin{equation}
\opSix_{1}=\frac{1}{\bare{g}^2}\tr([D_\mu, F_{\nu\rho}]\,[D_\mu, F_{\nu\rho}])
\,,\quad
\opSix_{2}=\frac{1}{\bare{g}^2}\sum\limits_{\mu}\tr([D_\mu, F_{\mu\nu}]\,[D_\mu, F_{\mu\nu}])\,.
\end{equation}
The extension to full QCD with massless quarks at $\ord(a^2)$ then introduces an additional fermion bilinear compatible with chiral symmetry to the minimal on-shell basis after applying the EOMs and making use of integration by parts on the basis listed in~\cite{Sheikholeslami:1985ij}
\begin{equation}
\opSix_3=\sum_\mu\bar\Psi\gamma_\mu D_\mu^3\Psi.
\end{equation}
Both operators $\opSix_{2}$ and $\opSix_3$ break \cubicSymm{} symmetry.
Therefore their mixing under renormalisation will be fairly restricted as they cannot mix into any operator invariant under \cubicSymm{} symmetry -- assuming a regulator preserving this symmetry.
At mass-dimension~6 also 4-fermion operators contribute.
For Ginsparg-Wilson quarks only those compatible with chiral symmetry are allowed
\begin{align}
\opSix_4&=\bare{g}^2(\bar\Psi\gamma_\mu\Psi)^2,& \opSix_5&=\bare{g}^2(\bar\Psi\gamma_\mu\gamma_5\Psi)^2,\nonumber\\
\opSix_6&=\bare{g}^2(\bar\Psi\gamma_\mu T^a\Psi)^2,& \opSix_7&=\bare{g}^2(\bar\Psi\gamma_\mu\gamma_5T^a\Psi)^2,\label{eq:4fermionChiral}
\end{align}
while Wilson quarks also require the inclusion of the chiral-symmetry violating 4-fermion operators
\begin{align}
\opSix_8&=\bare{g}^2(\bar\Psi\Psi)^2,& \opSix_9&=\bare{g}^2(\bar\Psi\gamma_5\Psi)^2,& \opSix_{10}&=\bare{g}^2(\bar\Psi\sigma_{\mu\nu}\Psi)^2,\nonumber\\
\opSix_{11}&=\bare{g}^2(\bar\Psi T^a\Psi)^2,& \opSix_{12}&=\bare{g}^2(\bar\Psi\gamma_5T^a\Psi)^2,& \opSix_{13}&=\bare{g}^2(\bar\Psi\sigma_{\mu\nu}T^a\Psi)^2.\label{eq:4fermionNonChiral}
\end{align}
\end{subequations}
We choose here the 4-fermion operator basis such that the flavour, colour algebra and spinor indices are contracted within $\bar{\Psi}\ldots\Psi$ pairs.
All other ways to contract the indices compatible with the symmetry constraints are related to the basis chosen here through the following identities.
Firstly through Fierz identities\footnote{Since we will use dimensional regularisation the Fierz identities are actually only correct up to Evanescent operators~\cite{Buras:1989xd} vanishing in four dimensions.
This ensures that such operators affect the renormalisation only beyond 1-loop.}
\begin{equation}
\begin{pmatrix}
\bar{\psi}_1\psi_2\bar\psi_3\psi_4\\[2pt]
\bar{\psi}_1\gamma_\mu\psi_2\bar\psi_3\gamma_\mu\psi_4\\[2pt]
\bar{\psi}_1\sigma_{\mu\nu}\psi_2\bar\psi_3\sigma_{\mu\nu}\psi_4\\[2pt]
\bar{\psi}_1\gamma_\mu\gamma_5\psi_2\bar\psi_3\gamma_\mu\gamma_5\psi_4\\[2pt]
\bar{\psi}_1\gamma_5\psi_2\bar\psi_3\gamma_5\psi_4
\end{pmatrix}=\frac{1}{8}\begin{pmatrix}
-2  &-2 &-1 & 2 &-2 \\[2pt]
-8  & 4 & 0 & 4 & 8 \\[2pt]
-24 & 0 & 4 & 0 &-24\\[2pt]
 8  & 4 & 0 & 4 &-8 \\[2pt]
-2  & 2 &-1 &-2 &-2
\end{pmatrix}
\begin{pmatrix}
\bar{\psi}_1\psi_4\bar\psi_3\psi_2\\[2pt]
\bar{\psi}_1\gamma_\mu\psi_4\bar\psi_3\gamma_\mu\psi_2\\[2pt]
\bar{\psi}_1\sigma_{\mu\nu}\psi_4\bar\psi_3\sigma_{\mu\nu}\psi_2\\[2pt]
\bar{\psi}_1\gamma_\mu\gamma_5\psi_4\bar\psi_3\gamma_\mu\gamma_5\psi_2\\[2pt]
\bar{\psi}_1\gamma_5\psi_4\bar\psi_3\gamma_5\psi_2
\end{pmatrix},\label{eq:4fermionFierz}\\
\end{equation}
where $\psi_n$ is a quark field with an arbitrary combination of flavour and colour index, while the spinor indices are contracted within the quark anti-quark pair.
Secondly through identities from the $\su(\Nc)$ algebra
\begin{align}
\bar{\psi}_A\Gamma_{\{\mu\}}T^a_{AB}\psi_B\bar{\eta}_C\Gamma_{\{\mu\}}T^a_{CD}\eta_D&=\frac{\TF}{N}\bar{\psi}_A\Gamma_{\{\mu\}}\psi_A\bar{\eta}_B\Gamma_{\{\mu\}}\eta_B-\TF\bar{\psi}_A\Gamma_{\{\mu\}}\psi_B\bar{\eta}_B\Gamma_{\{\mu\}}\eta_A,\label{eq:4fermionFierzSUN1}\\
\bar{\psi}_A\Gamma_{\{\mu\}}T^a_{AB}\eta_B\bar{\eta}_C\Gamma_{\{\mu\}}T^a_{CD}\psi_D&=\frac{\TF}{N}\bar{\psi}_A\Gamma_{\{\mu\}}\eta_A\bar{\eta}_B\Gamma_{\{\mu\}}\psi_B-\TF\bar{\psi}_A\Gamma_{\{\mu\}}\eta_B\bar{\eta}_B\Gamma_{\{\mu\}}\psi_A\label{eq:4fermionFierzSUN2},
\end{align}
where $\psi$ and $\eta$ indicate different flavours, $\Gamma_{\{\mu\}}$ denotes a matrix of the Dirac algebra with all indices $\{\mu\}$ contracted with the second  $\Gamma_{\{\mu\}}$ and $A$, $B$, $C$, $D$ are contracted colour indices.
This particular choice for the basis is identical to the one in~\cite{Bar:2003mh} and, as shown there, equivalent to the choice in~\cite{Sheikholeslami:1985ij}.
We choose to prepend an additional factor of $\bare g^2$ to each 4-fermion operator motivated by the gluonic EOM~\eqref{eq:EOMs} as well as the leading order of any kind of tree-graph leading to a 4-fermion interaction in lattice QCD.
The latter happens due to the absence of terms with more than one quark-anti-quark pair in the classical expansion of the lattice action in the lattice spacing $a$ as discussed in~\cite{Sheikholeslami:1985ij}, i.e.~the tree-level coefficients of 4-fermion operators without the factor $\bare g^2$ would vanish anyway.

\subsection{Massive operator basis}\label{sec:massiveBasis}
Unless one is interested in massless quarks or quarks at very small quark masses, like \mbox{up-,} down- or strange-quarks, operators carrying explicit powers of quark masses should be taken into account because they will no longer be suppressed.
If a hadronic renormalisation scheme is used on the lattice rather than a mass-independent scheme, such contributions can be ignored.
For the general massive case one finds at mass-dimension~5 the additional linearly independent operators
\begin{align}
\opFive_2&=\frac{\tr(M)}{\Nf}\frac{1}{\bare{g}^2}\tr(F_{\mu\nu}F_{\mu\nu}),&
\opFive_3&=\bar{\Psi}M^2\Psi,\nonumber\\
\opFive_4&=\frac{\tr(M)}{\Nf}\bar{\Psi}M\Psi,&
\opFive_5&=\frac{\tr(M^2)}{\Nf}\bar{\Psi}\Psi,\vphantom{\frac{1}{\bare{g}^2}}\nonumber\\
\opFive_6&=\frac{\tr(M)^2}{\Nf^2}\bar{\Psi}\Psi,
\end{align}
where the normalisation with $\Nf$ is chosen to ensure that in the mass-degenerate case all operators carry the same prefactor.
Similarly, the basis of mass-dimension~6 operators requires the inclusion of the larger number of additional linearly independent massive operators
\begin{align}
\opSix_{14}&=\frac{i}{4}\bar{\Psi}M\sigma_{\mu\nu}F_{\mu\nu}\Psi,&
\opSix_{15}&=\frac{\tr(M^2)}{\Nf}\frac{1}{\bare{g}^2}\tr(F_{\mu\nu}F_{\mu\nu}),\nonumber\\
\opSix_{16}&=\bar{\Psi}M^3\Psi,&
\opSix_{17}&=\frac{\tr(M^2)}{\Nf}\bar{\Psi}M\Psi,\nonumber\\
\opSix_{18}&=\frac{i\tr(M)}{4\Nf}\bar{\Psi}\sigma_{\mu\nu}F_{\mu\nu}\Psi,&
\opSix_{19}&=\frac{\tr(M)^2}{\Nf^2}\frac{1}{\bare{g}^2}\tr(F_{\mu\nu}F_{\mu\nu}),\nonumber\\
\opSix_{20}&=\frac{\tr(M)}{\Nf}\bar{\Psi}M^2\Psi,&
\opSix_{21}&=\frac{\tr(M)^2}{\Nf^2}\bar{\Psi}M\Psi,\nonumber\\
\opSix_{22}&=\frac{\tr(M^3)}{\Nf}\bar{\Psi}\Psi,&
\opSix_{23}&=\frac{\tr(M^2)\tr(M)}{\Nf^2}\bar{\Psi}\Psi,\nonumber\\
\opSix_{24}&=\frac{\tr(M)^3}{\Nf^3}\bar{\Psi}\Psi.&
\end{align}
The way an explicit mass term added to the lattice action reduces the flavour symmetries of Ginsparg-Wilson fermions to $\mathrm{U}(1)_\mathrm{V}^{\Nf}$ implies that only the operators $\opSix_{14-17}$ are allowed to contribute\footnote{We thank the referee for pointing out this simplification.
Due to the degenerate anomalous dimensions, this will have no impact on the spectrum of powers in $\gbar^2$ found in section~\ref{sec:matching}.}.
To see how this simplification arises, we first rewrite the mass term as
\begin{equation}
\bar{\Psi}M\Psi = \bar{\Psi}_RM\Psi_L+\bar{\Psi}_LM^\dagger\Psi_R\,,
\end{equation}
where the subscripts indicate left- and right-handed quarks.
Promoting the mass matrix $M$ to a spurionic field that transforms according to
\begin{equation}
M\rightarrow RML^\dagger,\quad L\in\mathrm{SU}(\Nf)_\mathrm{L},\quad R\in\mathrm{SU}(\Nf)_\mathrm{R},
\end{equation}
ensures that the action stays invariant under the flavour rotations
\begin{equation}
\bar{\Psi}_L\rightarrow \bar{\Psi}_LL^\dagger,\quad \bar{\Psi}_R\rightarrow \bar{\Psi}_RR^\dagger,\quad \Psi_L\rightarrow L\Psi_L,\quad \Psi_R\rightarrow R\Psi_R\,.
\end{equation}
For higher powers of the mass matrix in our SymEFT basis, we replace iteratively all occurrences of $M^2\rightarrow MM^\dagger$ from the left.
Requiring invariance under the spurionic symmetry transformation then indeed leaves only the $\opSix_{14-17}$ as stated in the beginning.
For Wilson quarks however, all massive mass-dimension~6 operators listed here are needed, because chiral symmetry was already explicitly broken by the lattice discretisation of the massless theory.

\section{Wilson quarks with a chiral twist}
When discussing Wilson quarks also Wilson quarks with a chirally twisted mass term~\cite{Frezzotti:1999vv,Frezzotti:2000nk,Frezzotti:2005gi} known as twisted mass QCD~(tmQCD) come to mind.
The lattice fermion action reads
\begin{equation}
S_\mathrm{tw}=a^4\sum_x\bar{\chi}(x)\Dtw\chi(x),
\end{equation}
where $\chi$ are the chirally twisted flavours and the Dirac operator is defined as
\begin{equation}
\Dtw=\frac{1}{2}\left\{\gamma_\mu(\nabla_\mu^*+\nabla_\mu)-a\nabla_\mu^*\nabla_\mu\right\}+\mq+i\gamma_5\tau^3\muq+a\frac{\csw(g_0^2)}{4}i\sigma_{\mu\nu}\hat{F}_{\mu\nu}.\label{eq:DtmQCD}
\end{equation}
Here $\tau^3$ is the Pauli matrix $\tau^3=\diag(1,-1)$ acting in flavour space and $\muq$ is the twisted mass parameter.
The generalisation to mass-non-degenerate flavour doublets exists~\cite{Frezzotti:2003xj}, but lies beyond the scope of this paper.
The reasoning used here should carry over, but it relies more heavily on spurionic symmetry arguments.
Also the overall renormalisation arguments for extracting the anomalous dimension matrix must be revisited.
Therefore we limit considerations here to the case of multiple mass-degenerate flavour doublets.

Due to the twisted mass term, invariance under parity transformation and time reversal are broken.
Instead the theory of the chirally twisted flavours~$\chi$ (both continuum and lattice) is invariant under~\cite{Frezzotti:1999vv} modified parity
\begin{equation}
\bar{\chi}(x_0,\vecx)\rightarrow -i\bar\chi(x_0,-\vecx) \gamma_0\tau^j,\quad\chi(x_0,\vecx)\rightarrow i\gamma_0\tau^j\chi(x_0,-\vecx)
\end{equation}
and modified time reversal
\begin{equation}
\bar{\chi}(x_0,\vecx)\rightarrow -i\bar\chi(-x_0,\vecx) \gamma_5\gamma_0\tau^j,\quad\chi(x_0,\vecx)\rightarrow i\gamma_0\gamma_5\tau^j\chi(-x_0,\vecx),
\end{equation}
where $j=1,2$, as well as the spurionic symmetry transformations
\begin{align}
\bar{\chi}(x_0,\vecx)&\rightarrow \bar\chi(x_0,-\vecx) \gamma_0,&\chi(x_0,\vecx)&\rightarrow \gamma_0\chi(x_0,-\vecx),&\muq&\rightarrow-\muq\label{eq:spurionic1}\\
\bar{\chi}(x_0,\vecx)&\rightarrow \bar\chi(-x_0,\vecx) \gamma_5\gamma_0,&\chi(x_0,\vecx)&\rightarrow \gamma_0\gamma_5\chi(-x_0,\vecx),&\muq&\rightarrow-\muq.\label{eq:spurionic2}
\end{align}
This does not have any impact on the massless operator basis, which is of course unchanged as the massless cases of Wilson QCD and tmQCD are identical.
For the massive basis the spurionic symmetry transformations limit the occurrence of powers of $\muq$ severely.

In the continuum theory and thus in our SymEFT twisted ($\chi$) and untwisted ($\Psi$) flavours are connected via the substitution
\begin{equation}
\bar\Psi= \bar\chi\exp(i\omega \gamma_5\tau^3/2),\quad \Psi=\exp(i\omega\gamma_5\tau^3/2)\chi,\quad \tan(\omega)=\frac{\muq}{\mq},\label{eq:twistSubst}
\end{equation}
as long as a regularisation is chosen, that is invariant under chiral rotations.
This connection enables us to infer the anomalous dimensions of twisted operators from the untwisted operators due to, see e.g.~\cite{Frezzotti:2001ea},
\begin{align}
\mvl{\Sc[i]\op[\bar{\Psi},\Psi] O_\mathrm{ext}}_\mathrm{QCD}&=Z^\op_{ij}\mvl{\op_j[\bar{\Psi},\Psi] O_\mathrm{ext}}_\mathrm{QCD}\nonumber\\
&\equiv Z^\op_{ij}\mvl{\op_j[\bar\chi\exp(i\omega \gamma_5\tau^3/2),\exp(i\omega\gamma_5\tau^3/2)\chi]O_\mathrm{ext}}_\mathrm{tmQCD},\label{eq:(non)twistedOps}
\end{align}
where the subscripts QCD and tmQCD denote the choice of the mass term in the action with flavours denoted as $\Psi$ and $\chi$ respectively.
Here $O_\mathrm{ext}$ is assumed to be invariant under \eq{eq:twistSubst}.
This equivalence holds in case a mass-independent multiplicative renormalisation scheme~$\mathcal{S}$ is used, which ensures that the mixing matrix is independent of the twist angle~$\omega$.

In contrast to the continuum action, the Wilson term in the lattice action \eq{eq:DtmQCD} obstructs the transformation \eq{eq:twistSubst}.
This requires the inclusion of additional operators to our SymEFT.
These additional operators relevant up to mass-dimension~6 are
\begin{align}
\bar{\chi}\chi,&& i\bar{\chi}\sigma_{\mu\nu}F_{\mu\nu}\chi,
\end{align}
dressed by an even power of the twisted mass parameter $\muq$ to comply with the spurionic symmetry transformations in~\eqs{eq:spurionic1} and \eqref{eq:spurionic2}.
These operators are accompanied by the chirally twisted versions of $\bar{\Psi}\Psi$ and $i\bar{\Psi}\sigma_{\mu\nu}\Psi$, namely
\begin{align}
i\bar{\chi}\gamma_5\tau^3\chi,&&\bar{\chi}\sigma_{\mu\nu}\tilde{F}_{\mu\nu}\tau^3\chi,\label{eq:twistedOld}
\end{align}
where $\tilde{F}_{\mu\nu}=\varepsilon_{\mu\nu\rho\sigma}F_{\rho\sigma}/2$ is the dual field strength tensor.
Both operators in \eq{eq:twistedOld} must be dressed by an odd power in $\muq$, to comply with the spurionic symmetry transformations in~\eqs{eq:spurionic1} and \eqref{eq:spurionic2}.
Any new 4-fermion operators at mass-dimension~6 will require an odd power in $\muq$, such that they eventually contribute only at mass-dimension~7, i.e.~$\ord(a^3)$, or beyond.

While the reasoning in \eq{eq:(non)twistedOps} is sufficient to fix the renormalisation of the massless operator basis and some mixing contributions of the operators in \eq{eq:twistedOld}, we need additional input to take also the mixing of the additional massive operators into account.
Since the computations we report here have been performed at zero twist angle we lack any information about massive mixing contributions involving $\muq$.
The only information we have for the massive operator basis are the diagonal entries rather than the full mixing.

A special property of twisted mass QCD is automatic $\ord(a)$ improvement at maximal twist, i.e.~$\omega=\pi/2$.
This is due to the additional discrete symmetry of the continuum theory, which reads at this twist angle~\cite{Aoki:2006gh,Sint:2007ug}
\begin{equation}
T_1:\quad \bar{\chi}\rightarrow i\bar{\chi}\gamma_5\tau^1,\quad\chi\rightarrow i\gamma_5\tau^1\chi.\label{eq:T1symmTransf}
\end{equation}
While being a continuum symmetry, it is again explicitly broken on the lattice by the Wilson term, which allows for additional $\ord(a)$ terms.
Following the lines of~\cite{Aoki:2006gh} any operator can be split into a $T_1$-even and $T_1$-odd part, i.e., parts having eigenvalues $\pm1$ under the transformation in \eq{eq:T1symmTransf} respectively.
This carries over to $n$-point functions of operators, which can then be split into a $T_1$-even and $T_1$-odd part as well, where the $T_1$-odd part vanishes by construction.
At maximal twist all mass-dimension~5 operators parametrising the $\ord(a)$ corrections are $T_1$-odd.
Thus any $\ord(a)$ lattice artifacts vanish for quantities that are themselves $T_1$-even.
However this does not imply that the matching coefficients $\omega_i$ of the mass-dimension~5 operators are zero.
Instead these operators become relevant at $\ord(a^2)$ through $T_1$-even double operator insertions but can in principle be removed through non-perturbative $\ord(a)$ improvement~\cite{Frezzotti:2005gi} identically to untwisted Wilson quarks~\cite{Luscher:1996ug}.
Without non-perturbative improvement these double operator insertions also give rise to contact terms in the SymEFT expansion to $\ord(a^2)$.

\def\Psiq{\check\Psi}
\def\Psiqbar{\check{\bar\Psi}}

\section{Partially quenched QCD and mixed actions}
With the results for full QCD at hand we can also consider the special case of partially quenched QCD, where the determinant of the lattice Dirac operator has been dropped for $\Nb$ flavours.
In perturbation theory this leads to discarding any closed fermion loops, while fermion lines contracted with local fields are kept.
Conceptually this can be implemented by introducing additional bosonic fields $\Phi=(\phi_1,\ldots,\phi_{\Nb})^\mathrm{T}$ into the lattice fermion action
\begin{equation}
S_\mathrm{F}=a^4\sum_x\left\{\bar\Psi(x)\hat D\Psi(x)+\Phi^\dagger(x)\hat D\Phi(x)\right\}\equiv a^4\sum_x\Psiqbar(x)\hat D\Psiq(x)
\end{equation}
such that closed fermion loops of the quenched flavours cancel out with closed boson loops, see e.g.~\cite{Morel:1987xk,Labrenz:1996jy,Bar:2005tu}.
Here we also introduced the short-hand
\begin{equation}
\Psiq=\begin{pmatrix}
\Psi \\
\Phi
\end{pmatrix}=(\psi_1,\ldots,\psi_{\Nf},\phi_1,\ldots,\phi_{\Nb})^\mathrm{T}.
\end{equation}

Assuming that the same lattice Dirac operator $\hat D$ is used for both quenched and unquenched flavours the underlying flavour symmetries are modified to the $\UN[\Nf\mid\Nb]{V}$ and $\SUN[\Nf\mid\Nb]{L}\times \SUN[\Nf\mid\Nb]{R}\times\UN[1]{V}$ graded symmetry for massless Wilson and Ginsparg-Wilson quarks respectively.
For some details on Lie superalgebras see e.g.~\cite[p.~9ff.]{Freund:1986ws}.
We only need to understand the constraint the graded flavour symmetry transformation
\begin{align}
\Psiqbar&=\begin{pmatrix}\bar\Psi & \Phi^\dagger\end{pmatrix}\rightarrow \begin{pmatrix}\bar\Psi & \Phi^\dagger\end{pmatrix}
\begin{pmatrix}
\tilde A & \tilde C \\
\tilde B & \tilde D
\end{pmatrix},\nonumber\\
\Psiq&=\begin{pmatrix}
\Psi \\
\Phi
\end{pmatrix}\rightarrow
\begin{pmatrix}
A & B \\
C & D
\end{pmatrix}\begin{pmatrix}
\Psi \\
\Phi
\end{pmatrix}
\end{align}
imposes on the minimal operator basis of our effective action, where
\begin{align}
\tilde{A}A+\tilde C C=\mathbb{1}_{\Nf\times\Nf},&&
\tilde{B}B+\tilde D D=\mathbb{1}_{\Nb\times\Nb},\nonumber\\
\tilde A B+\tilde C D=\mathbb{0}_{\Nf\times\Nb},&&\tilde B A+\tilde D C=\mathbb{0}_{\Nb\times\Nf}.
\end{align}
The generalisation necessary for our operators containing fermion fields can be inferred from the transformation of fermion bilinears
\begin{align}
&\bar\Psi\Gamma\Psi\rightarrow \bar\Psi'\Gamma\Psi'=\bar\Psi\tilde A A\Gamma\Psi+\Phi^\dagger\tilde B B\Gamma\Phi+\Phi^\dagger\tilde B A\Gamma\Psi+\bar\Psi\tilde A B\Gamma\Phi\neq\bar\Psi\Gamma\Psi,\nonumber\\
&\text{if }(B\neq0)\vee(\tilde{B}\neq0)\vee(\tilde{C}C\neq0),
\end{align}
where $\Gamma$ is flavour diagonal and, in the case of Ginsparg-Wilson fermions, preserves chiral symmetry.
An analogous transformation can be given for bosonic bilinears.
The special case of $B=\tilde C=0=C=\tilde B,$ realises separate rotations in fermionic and bosonic flavour space but this corresponds only to a subset of the full graded flavour symmetry.
We thus immediately see that fermion bilinears and 4-fermion operators must be generalised as $\Psi\rightarrow\Psiq$ for the partially quenched case.
This also means that there are no operators needed besides those in~\eqs{eq:minBasis} with the substitution to partially quenched flavours applied.

For the sake of argument mixed actions, i.e.~different choices for the Dirac operator of sea ($\hat{D}_\mathrm{S}$) and valence ($\hat{D}_\mathrm{V}$) quarks, can be written as a partially quenched theory of $\Nf$ unquenched and $\Nb$ quenched flavours~\cite{Bar:2003mh}
\begin{equation}
S_\mathrm{F}=a^4\sum_x\left\{\bar{\Psi}_\mathrm{S}(x)\hat{D}_\mathrm{S}\Psi_\mathrm{S}(x)+\Psiqbar_\mathrm{V}(x)\hat{D}_\mathrm{V}\Psiq_\mathrm{V}(x)\right\}.
\end{equation}
The quenched flavours play the role of the valence quarks, while the unquenched flavours are the sea quarks.
Due to the different discretisations of the Dirac operators $\hat{D}_\mathrm{S}$ and $\hat{D}_\mathrm{V}$, the flavour symmetries are more complicated as separate flavour rotation symmetries are restricted to the quenched or unquenched flavours respectively.
Thus we expect the following minimal operator basis
\begin{align}
\opSix_{1}&=\frac{1}{g_0^2}\tr([D_\mu, F_{\nu\rho}]\,[D_\mu, F_{\nu\rho}])
\,,&
\opSix_{2}&=\frac{1}{g_0^2}\sum\limits_{\mu}\tr([D_\mu, F_{\mu\nu}]\,[D_\mu, F_{\mu\nu}])\,,\nonumber\\
\opSix_{3;\mathrm{S}}&=\sum\limits_{\mu}\bar{\Psi}_\mathrm{S}\gamma_\mu D_\mu^3\Psi_\mathrm{S},&\opSix_{i;\mathrm{S}}&=g_0^2(\bar{\Psi}_\mathrm{S}\Gamma_{i;\mathrm{S}}\Psi_\mathrm{S})^2,\nonumber\\
\opSix_{3;\mathrm{V}}&=\sum\limits_{\mu}\Psiqbar_\mathrm{V}\gamma_\mu D_\mu^3\Psiq_\mathrm{V},&\opSix_{i;\mathrm{V}}&=g_0^2(\Psiqbar_\mathrm{V}\Gamma_{i;\mathrm{V}}\Psiq_\mathrm{V})^2,\vphantom{\sum\limits_\mu}\nonumber\\
\opSix_{i;\mathrm{SV}}&=g_0^2(\bar\Psi_\mathrm{S}\Gamma_{i;\mathrm{SV}}\Psi_\mathrm{S})(\Psiqbar_\mathrm{V}\Gamma_{i;\mathrm{SV}}\Psiq_\mathrm{V}),\label{eq:mixedActionBasis}
\end{align}
where $\Gamma_{i;\mathrm{S}/\mathrm{V}}$ are substitutes for all Dirac matrices allowed by the respective symmetry constraints with and without additional colour algebra generator $T^a$.
The same holds for $\Gamma_{i;\mathrm{SV}}$, but here the more restrictive symmetry constraints of either the quenched or the unquenched flavours decide, which Dirac matrices are allowed.
In the massive case the minimal basis is enlarged accordingly.

Instead of sea and valence quarks we could just as well introduce different discretisations for different sets of flavours.
In this case the reasoning still remains the same just without the quenching.

\section{One-loop computation of the anomalous dimension matrix}
\label{sec:AD}
\begin{figure}
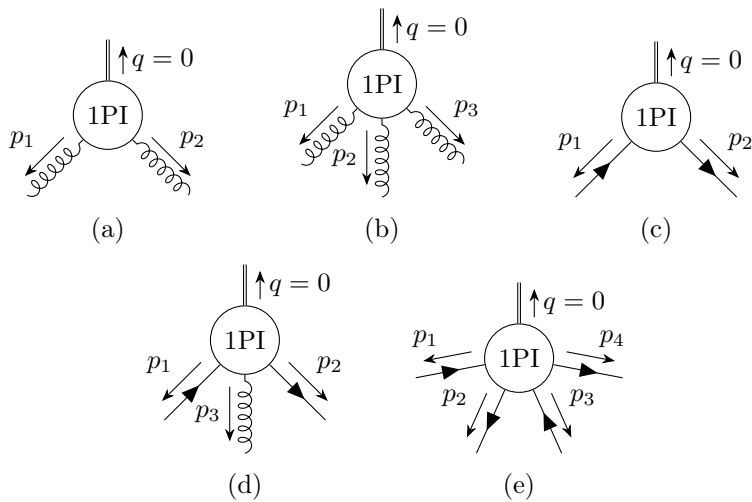

\centering
\begin{subfigure}[t]{0.232\textwidth}\centering
\includegraphics[]{\imgPath A2O.pdf}
\caption{}
\label{fig:B2O}
\end{subfigure}
\begin{subfigure}[t]{0.232\textwidth}\centering
\includegraphics[]{\imgPath A3O.pdf}
\caption{}
\label{fig:B3O}
\end{subfigure}
\begin{subfigure}[t]{0.232\textwidth}\centering
\includegraphics[]{\imgPath F2O.pdf}
\caption{}
\label{fig:F2O}
\end{subfigure}
\vskip5pt
\begin{subfigure}[t]{0.232\textwidth}\centering
\includegraphics[]{\imgPath F2AO.pdf}
\caption{}
\label{fig:F2BO}
\end{subfigure}
\begin{subfigure}[t]{0.232\textwidth}\centering
\includegraphics[]{\imgPath F4O.pdf}
\caption{}
\label{fig:F4O}
\end{subfigure}
\caption{Graphical representation of all 1PI graphs of fundamental quark fields or background fields with insertion of an operator $\op$ or $\opE$ considered for the renormalisation of the basis.
Here the double line indicates the momentum contribution of the inserted operator, which may also be seen as an additional leg and is set to zero momentum.
The wiggly lines are external background fields and the straight lines carrying arrows are quarks.
The graph (e) is only needed at mass-dimension~6 to include 4-fermion operators.}
\label{fig:1PIgraphs}
\end{figure}

To renormalise the full QCD on-shell basis perturbatively to 1-loop order, we make use of the background field method~\cite{DeWitt:1967ub,KlubergStern:1974xv,Abbott:1980hw,Luscher:1995vs}.
There, a smooth classical background field $B_\mu(x)$ is introduced as 
\begin{equation}
A_\mu(x)= B_\mu(x) + g_0 Q_\mu(x) \,,
\end{equation}
splitting the quantum field $A_\mu(x)$ into the background field and the quantum fluctuations~$Q_\mu(x)$.
Additionally the background field gauge is chosen by adding the gauge-fixing term
\begin{equation}
\Lbf(B,Q)=-\lambda_0\tr([D_\mu(B),Q_\mu] [D_\nu(B),Q_\nu])
\end{equation}
to the continuum Lagrangian with bare gauge-fixing parameter~$\lambda_0$ alongside a Faddeev-Popov term~\cite{Faddeev:1967}.
Due to the background field gauge, only the fields $Q_\mu$ are gauge-fixed, while the background field itself remains unaffected.
Then one-particle-irreducible~(1PI) graphs with only background fields and quarks as legs, see also \fig{fig:1PIgraphs}, transform manifestly gauge-covariant under gauge-transformations of the background field.
This ensures absence of contributions from non-gauge-invariant operators mixing into the gauge-invariant ones, which are in principle allowed due to working in the gauge-fixed theory~\cite{Joglekar:1975nu,Collins:1994ee}.
Such mixing contributions are of course irrelevant when being interested in gauge-invariant observables.
However, we cannot ignore contributions from gauge-invariant operators~$\opE$ vanishing by the equations of motion.
Again such contributions vanish on-shell and the mixing of EOM vanishing operators under renormalisation is therefore restricted as
\begin{equation}
\begin{pmatrix}
\op \\[6pt]
\opE
\end{pmatrix}_\mathrm{R}=\begin{pmatrix}
Z & A^{\op \opE} \\[6pt]
0 & Z^\opE
\end{pmatrix}\begin{pmatrix}
\op \\[6pt]
\opE
\end{pmatrix},
\end{equation}
where $Z$ is the desired on-shell mixing matrix of our operator basis.
For compactness we omitted here all operator indices, i.e.~$Z$, $A^{\op\opE}$ and $Z^\opE$ are matrices themselves.
We thus need an additional minimal basis of EOM vanishing operators
\begin{align}
\opFourE_1&=\bar{\Psi}[\gamma_\mu D_\mu+ M]\Psi\vphantom{\frac{1}{\bare g^2}},
&&&\opFiveE_1&=\bar{\Psi}[\gamma_\mu D_\mu+ M]^2\Psi\vphantom{\frac{1}{\bare g^2}},\nonumber\\
\opSixE_1&=\bar{\Psi}[\gamma_\mu D_\mu+ M]^3\Psi\vphantom{\frac{1}{\bare g^2}},&&&
\opSixE_2&=\bar{\Psi}\gamma_\mu [D_\nu, F_{\nu\mu}]\Psi-\bare g^2(\bar\Psi \gamma_\mu T^a\Psi)^2,\vphantom{\frac{1}{\bare g^2}}\nonumber\\
\opSixE_3&=\frac{1}{\bare g^2}\mathrlap{\tr([D_\mu, F_{\mu\nu}][D_\rho, F_{\rho\nu}])+\frac{1}{2}\bar\Psi \gamma_\mu [D_\nu, F_{\nu\mu}]\Psi,\vphantom{\frac{1}{\bare g^2}}}
\end{align}
where the lower dimensional ones may carry additional powers of masses.

With these prerequisites we can now renormalise our minimal operator basis by computing the 1PI graphs with single operator insertion as depicted in \fig{fig:1PIgraphs} using dimensional regularisation~($D=4-2\epsilon$) and renormalising any divergences in the modified minimal subtraction~($\MSbar$) scheme~\cite{tHooft:1972tcz,tHooft:1973mfk,Bardeen:1978yd}.

Working in a mass-independent renormalisation scheme like $\MSbar$ ensures that the 1-loop anomalous dimensions are simply related to counterterms renormalising ultraviolet divergences.
We then use the by now common trick called \textit{infrared rearrangement}, see e.g.~\cite{Misiak:1994zw,Chetyrkin:1997fm,Luthe:2017ttc}.
This enables us to separate the UV-divergent part of a 1-loop momentum integral from UV-finite but potentially IR divergent parts through the exact relation~\cite{Misiak:1994zw,Chetyrkin:1997fm}
\begin{equation}
\frac{1}{(k+p)^2+m^2}=\frac{1}{k^2+\Omega}-\frac{2kp+p^2+m^2-\Omega}{[k^2+\Omega][(k+p)^2+m^2]}\,.\label{eq:IRrearrangement}
\end{equation}
Here $k$ is the loop momentum, $p$ is an external momentum, $m$ is a mass and we choose $\Omega>0$ as a mass-scale.
The second term in \eq{eq:IRrearrangement} on the r.h.s.~is one power less UV-divergent.
Iterating this step brings all UV-divergent terms into the standard form $\int\rmd^Dk\,[k^2+\Omega]^{-n}k_{\mu_1}\ldots k_{\mu_l}$, while all UV-finite terms can eventually be dropped as they do not contribute to the 1-loop anomalous dimensions.
These steps have been implemented in \FORM/~\cite{Vermaseren:2000nd} to evaluate the various 1PI graphs at 1-loop\footnote{The extraction of the 1-loop anomalous dimensions has been automated to some extent via a \texttt{Makefile} based around \texttt{FORM} scripts, a \texttt{QGRAF} model file and some \texttt{Python} scripts.
For the final renormalisation step some Mathematica routines were used.
All of this can be accessed from~\url{https://github.com/nikolai-husung/Symanzik-QCD-workflow}.}\kern-0.5em.
We also derived the necessary Feynman rules of the various operators in~\FORM/ and checked that we reproduce the usual Feynman rules for full QCD in the background field gauge.

Instead of inserting the flavour singlet operators, we insert variants where the flavour is kept generic and we can then build the required set of operators from these building blocks.
As a consequence we are immediately able to extract the generalisations necessary for cases like partially quenched and mixed actions without too much effort.
All necessary graphs are obtained using \QGRAF/~\cite{Nogueira:1993,Nogueira:2006pq}.
The operator insertions are realised by formally introducing additional non-propagating fields $\varphi_i$ called ``anchor'' and adding $\sum_i\varphi_i\op_i$ to the Lagrangian.
These anchors correspond to the double lines in \fig{fig:1PIgraphs}.
For all $n$-point functions with a single operator insertion, except single flavour 4-fermion operators, setting
\begin{center}
\texttt{options = onepi;}
\end{center}
in \QGRAF/ is sufficient.
The single flavour 4-fermion operators are implemented by splitting the 4-fermion vertex into two 3-vertices connected with an additional ``mediator'' (here denoted by the dotted line) according to
\begin{align}
\raisebox{-.48\height}{\includegraphics{\imgPath fullOp}}&=\raisebox{-.48\height}{\includegraphics{\imgPath partialOp1}}
-\raisebox{-.48\height}{\includegraphics{\imgPath partialOp2}}\\[6pt]
&= 2\bare{g}^2(\delta_{AB}\delta_{CD}\Gamma_{ij}\Gamma_{kl}-\delta_{AD}\delta_{CB}\Gamma_{il}\Gamma_{kj})\label{eq:4fermFeynmanRule}\\
\raisebox{-.48\height}{\includegraphics{\imgPath mediatorOp}}&=\sqrt{2}g_0\Gamma_{ij}\delta_{AB}.
\end{align}
The ``mediator'' ensures that relative minuses due to anti-commutativity of fermions are taken care of\footnote{We do not claim that single flavoured 4-fermion operators would not work in \texttt{QGRAF} without this trick.
We just wanted to make sure that no surprises occur and that we do not accidentally introduce an additional relative minus.}\kern-0.5em~and that any spacetime indices encapsulated in $\Gamma$ are equal at the second vertex of this kind completing the 4-fermion operator.
For the single flavour 4-fermion operator setup we thus have to change the parameters for \QGRAF/ in this case to
\begin{center}
\texttt{options = ;\\
true = iprop[mediator,2,2];\\
true = bridge[quark,0,0];\\
true = bridge[gluon,0,0];}
\end{center}
This takes into account that cutting the propagator of the ``mediator'' cannot imply a reducible graph and thus must be included in the set of 1PI graphs.
In the same fashion one can also define $(\bar{q}\Gamma T^aq)^2$.
Eventually we are interested in the flavour singlets and, for the mixed action, in combinations of flavour singlets in two different flavour subsets.
Both can be easily obtained from the results for the flavoured operators.

Composite operators can mix according to the symmetry constraints, e.g.~operators compatible with $\SUN[\Nf]{L}\times\SUN[\Nf]{R}$ flavour symmetry can mix into operators with reduced $\SUN[\Nf]{V}$ flavour symmetry but not the other way around.
This mixing pattern is thus present in the mixing matrix $Z$ as well as in the closely related anomalous dimension matrix
\begin{equation}
\gamma^\op(\gbar)=\mu\frac{\D Z}{\D\mu}Z^{-1}=-\gbar^2\left\{\gamma_0^\op+\gamma_1^\op\gbar^2+\ord(\gbar^4)\right\}.
\end{equation}
Since we are only interested in the leading asymptotic dependence on the lattice spacing, we restrict considerations to the 1-loop coefficient $\gamma_0^\op$ of the anomalous dimension matrix, which has the form
\def\gluonic{\mathrm{g}}
\def\ch{\mathrm{L\mid R}}
\def\suv{\mathrm{V}}
\def\mm{\mathrm{m}}
\begin{equation}
\gamma_0^\op=\left(\begin{array}{ccc}
\gamma_0^{\ch}       & 0               & \gamma_0^{\ch,\mm} \\[4pt]
\gamma_0^{\suv,\ch}  & \gamma_0^{\suv} & \gamma_0^{\suv,\mm} \\[4pt]
0                    & 0                    & \gamma_0^{\mm}
\end{array}\right).\label{eq:mixingMatrixStructure}
\end{equation}
Notice that we hid the triangular structure of $\gamma_0^\op$ for the sake of compatibility to the original numbering of the operator basis.
The superscripts introduced here indicate the following subsets of operators:
\begin{itemize}
\item{$\ch$:}\\
Includes operators which are invariant under separate flavour rotations for left- and right-handed quarks.
\item{$\suv$:}\\
Operators which are only invariant under joint flavour rotations of left- and right-handed quarks.
These operators are in general needed for massless Wilson quarks due to less restrictive flavour symmetry constraints.
\item{$\mm$:}\\
These operators are lower-dimensional operators multiplied by explicit powers of masses.
\end{itemize}

For the partially quenched case, where only one lattice discretisation is used for all flavours, it suffices to make the replacements 
\begin{equation}
\Nf\rightarrow(\Nf-\Nb),\quad \tr(M^n)\rightarrow\tr(M_\mathrm{f}^n)-\tr(M_\mathrm{b}^n),
\end{equation}
where $M_\mathrm{f}$ and $M_\mathrm{b}$ are the diagonal mass matrices for the fermionic and bosonic flavours respectively.
We checked explicitly that setting $\Nf=0$ indeed yields the fully quenched approximation also for the 4-fermion operators.
This case can be easily obtained by discarding any closed fermion loops.

As a cross-check of our renormalisation procedure we also compared our mixing matrix for the non-singlet 4-fermion operators at $\Nf=3$ with the results found in the literature~\cite{Jamin:1985su,Boito:2015joa} - we agree.
To do so we made use of the the identities \eqs{eq:4fermionFierz} -- \eqref{eq:4fermionFierzSUN2} to eliminate the redundant (up to Evanescent operators) single flavour 4-fermion operators via
\begin{equation}
	\begin{pmatrix}
		(\bar{q} T^aq)^2\\[6pt]
		(\bar{q} \gamma_5 T^aq)^2\\[6pt]
		(\bar{q} \gamma_\mu T^aq)^2\\[6pt]
		(\bar{q} \gamma_\mu\gamma_5T^aq)^2\\[6pt]
		(\bar{q} \sigma_{\mu\nu}T^aq)^2
	\end{pmatrix}=\underbrace{\frac{1}{16}
\begin{pmatrix}
 \frac{8}{\Nc}+2 & 2 & 2 & -2 & 1 \\[6pt]
 2 & \frac{8}{\Nc}+2 & -2 & 2 & 1 \\[6pt]
 8 & -8 & \frac{8}{\Nc}-4 & -4 & 0 \\[6pt]
 -8 & 8 & -4 & \frac{8}{\Nc}-4 & 0 \\[6pt]
 24 & 24 & 0 & 0 & \frac{8}{\Nc}-4 
\end{pmatrix}}_{\define\mathcal{F}}
	\begin{pmatrix}
                (\bar{q}q)^2\\[6pt]
                (\bar{q} \gamma_5 q)^2\\[6pt]
                (\bar{q} \gamma_\mu q)^2\\[6pt]
                (\bar{q} \gamma_\mu\gamma_5 q)^2\\[6pt]
                (\bar{q} \sigma_{\mu\nu} q)^2
        \end{pmatrix}\label{eq:FierzSingleFlavour}.
\end{equation}
We stress again that this equivalence does not hold in dimensional regularisation, but for the renormalisation at 1-loop order this subtlety is of no consequence as we can safely ignore any Evanescent operators to this loop order~\cite{Buras:1989xd}.

Via use of \eq{eq:FierzSingleFlavour} also the special case of Wilson quarks with only one flavour $q$ without quenching can be formally obtained by the change of basis
\begin{equation}
\bare{g}^2(\bar{q}\Gamma_i T^a q)^2\rightarrow\bare{g}^2(\bar{q}\Gamma_i T^a q)^2-\mathcal{F}_{ij}\bare{g}^2(\bar{q}\Gamma_j q)^2
\end{equation}
introducing a set of Evanescent operators that mix only \textit{into} the other operators and not the other way around at $\Nf=1$.
Those Evanescent operators take the place of the ones of the form $\bare{g}^2(\bar{q}\Gamma_i T^a q)^2$ while we keep $\bare{g}^2(\bar{q}\Gamma_i q)^2$ in our minimal basis.
The desired mixing matrix can then be obtained from the subblock without the Evanescent operators.
Those renormalised Evanescent operators vanish to all orders in perturbation theory in the limit $D\rightarrow 4$, while their mixing contributions are needed to renormalise the remaining operators.
While $\Nf=1$ is an uncommon choice, the case of having e.g.~$\Nf=2+1$ or $\Nf=3+1$ with different lattice discretisations for the different flavour subsets is an interesting option\footnote{$\Nf=n+1$ denotes here different lattice discretisations for the $n$ flavours and the single flavour rather than different quark masses as is often done in the literature.}, that we want to include here as well for Wilson quarks in the single flavour subset.
Notice, that the case of $\Nf=n+1$ without quenching is special in the sense, that if we reached $\Nf=n+1$ by having more than one quark flavour in the second set of flavours but a sufficient number of bosonic flavours, the basis is a different one and we would instead need to include the generalisation of all 4-fermion operators to the partially quenched basis.

\subsection{Mass-dimension~5}
In contrast to the earlier paper~\cite{Husung:2019ytz} we will also include massive operators mixing into the ones without explicit mass-dependence.
Before discussing the operator mixing at mass-dimension~6 we thus give for completeness the full mixing matrix of the massive mass-dimension~5 operator basis
\begin{subequations}
\begingroup\allowdisplaybreaks
\begin{align}
(4\pi)^2\AD[\suv]{1}&=\frac{\Nc^2-5}{\Nc}\,,\\[6pt]
(4\pi)^2\AD[\suv,\mm]{1}&=\begin{pmatrix}
-2\Nf & 3\frac{1-\Nc^2}{\Nc} & 0 & 0 & 0
\end{pmatrix},\\[6pt]
(4\pi)^2\AD[\mm]{1}&=\begin{pmatrix}
-\frac{13}{3}\Nc-\frac{3}{\Nc}+\frac{4}{3}\Nf & 0 & 6\frac{\Nc^2-1}{\Nc}& 0 & 0 \\[6pt]
0 & 3\frac{\Nc^2-1}{\Nc} & 0 & 0 & 0 \\[6pt]
0 & 0 & 3\frac{\Nc^2-1}{\Nc} & 0 & 0 \\[6pt]
0 & 0 & 0 & 3\frac{\Nc^2-1}{\Nc} & 0 \\[6pt]
0 & 0 & 0 & 0 & 3\frac{\Nc^2-1}{\Nc}
\end{pmatrix},
\end{align}
\endgroup
\end{subequations}
where the corresponding minimal basis $\opFive_i$ is listed in \sects{sec:masslessBasis} and \ref{sec:massiveBasis}.
As mentioned earlier, there exists no massless operator that is invariant under separate flavour rotations for left- and right-handed quarks in our minimal basis at mass-dimension~5.

In anticipation of the discussion in section~\ref{sec:RGIs}, we give here the proper basis diagonal under 1-loop renormalisation
\begin{align}
\baseFive_1&=\opFive_1+\frac{\Nc\Nf}{1-\betaN\Nc+\Nc^2}\left\{\opFive_2+3\frac{1-\Nc^2}{1+\Nc^2}\opFive_4\right\}+\frac{3}{2}\frac{\Nc^2-1}{1+\Nc^2}\opFive_3,\nonumber\\
\baseFive_2&=\opFive_2+3\frac{1-\Nc^2}{\Nc\hat{b}_0}\opFive_4,\nonumber\\
\baseFive_{3-6}&=\opFive_{3-6}\,,\label{eq:basis[O]=5}
\end{align}
where $\betaN=(4\pi)^2b_0=\frac{11}{3}\Nc-\frac{2}{3}\Nf$ is related to the leading order coefficient $b_0$ of the $\beta$-function.
The associated normalised 1-loop coefficients $\ADhat[\base]{d}_{i}=\left(\gamma_0^\base\right)_i^{(d)}\!/(2b_0)$ of the anomalous dimension matrix are
\begin{equation}
\ADhat[\base]{1}_{1}=\frac{\Nc^2-5}{2\Nc\betaN},\quad\ADhat[\base]{1}_{2}=3\frac{\Nc^2-1}{2\Nc\betaN}-1\,,\quad\ADhat[\base]{1}_{3-6}=3\frac{\Nc^2-1}{2\Nc\betaN}\,.\label{eq:spectrum[O]=5}
\end{equation}
The coefficient $\ADhat[\base]{1}_1$ agrees with results found in the literature~\cite{Narison:1983}.

\subsection{Mass-dimension~6}
At mass-dimension~6 all 4-fermion operators contribute, which enlarges the mixing matrix quite a bit, and also the number of massive operators grows severely.
The diagonalisation of this operator basis is then not feasible for arbitrary values of $\Nf$ and $\Nc$ as will become clearer in section~\ref{sec:RGIs}.
We therefore give here only the subblocks of the non-diagonal mixing matrix.
For the massless case we find
\begin{subequations}\label{eq:mixingMassless1set}
\begingroup\allowdisplaybreaks
\begin{align}
(4\pi)^2\AD[\ch]{2}&=
\begin{pmatrix}
 A(\Nf) & B & C             \\[6pt]
 D(\Nf) & E & F(\Nf)        \\[6pt]
 \NULL_{4\times 2}      & \NULL_{4\times 1} & G(\betaN,\Nf)
\end{pmatrix},\\[6pt]
(4\pi)^2\AD[\suv]{2}&=J(\betaN),\\[6pt]
(4\pi)^2\AD[\suv,\ch]{2}&=\begin{pmatrix}
\NULL_{6\times 3} & H
\end{pmatrix},
\end{align}
\endgroup
\end{subequations}
where we introduced $\NULL_{m\times n}$ as the $m\times n$ matrix filled with zeros.
For reuse in the case of mixed actions in \sect{sec:ADmixed}, we also introduced the short-hands
\begin{subequations}\label{eq:mixingShorthands}
\begingroup\allowdisplaybreaks
\begin{align}
A(\Nf)&=
\begin{pmatrix}
 \frac{14 \Nc}{3}+\frac{4 \Nf}{3} & 0                                \\[6pt]
 -\frac{14 \Nc}{15}               & \frac{42 \Nc}{5}+\frac{4 \Nf}{3}
\end{pmatrix},\\
B&=\begin{pmatrix}
0 \\[6pt]
\frac{11}{30 \Nc}-\frac{11 \Nc}{30}
\end{pmatrix},\\[6pt]
C&=
\begin{pmatrix}
 0 & 3-\frac{3}{\Nc^2} & \frac{\Nc}{3}+\frac{4}{3 \Nc}        & \frac{12}{\Nc}-3 \Nc \\[6pt]
 0 & 1-\frac{1}{\Nc^2} & \frac{47 \Nc}{120}+\frac{19}{40 \Nc} & \frac{4}{\Nc}-\Nc
\end{pmatrix},\\[6pt]
D(\Nf)&=\begin{pmatrix}
 \frac{11 \Nf}{30} & -\frac{22 \Nf}{15}
\end{pmatrix},\\[6pt]
E&=\frac{157 \Nc}{30}-\frac{157}{30 \Nc},\\[6pt]
F(\Nf)&=\begin{pmatrix}
0 & \frac{1}{4}-\frac{1}{4 \Nc^2} & \frac{3 \Nc}{40}-\frac{13}{40 \Nc}-\frac{7 \Nf}{30} & \frac{1}{\Nc}-\frac{\Nc}{4}
\end{pmatrix}\\[6pt]
G(b,\Nf)&=\begin{pmatrix}
 2 b               & 0                 & -\frac{8}{3}                              & -12                  \\[6pt]
 0                 & 2 b               & -\frac{44}{3}                             & 0                    \\[6pt]
 0                 & \frac{3}{\Nc^2}-3 & 2 b-3 \Nc-\frac{4}{3 \Nc}+\frac{8 \Nf}{3} & 3 \Nc-\frac{12}{\Nc} \\[6pt]
 \frac{3}{\Nc^2}-3 & 0                 & 3 \Nc-\frac{40}{3 \Nc}                    & 2 b-3 \Nc
\end{pmatrix},\\[6pt]
H&=\begin{pmatrix}
 0 & 0 & \frac{4}{3}      & 0 \\[6pt]
 0 & 0 & -\frac{4}{3}     & 0 \\[6pt]
 0 & 0 & 0                & 0 \\[6pt]
 0 & 0 & \frac{2}{3 \Nc}  & 0 \\[6pt]
 0 & 0 & -\frac{2}{3 \Nc} & 0 \\[6pt]
 0 & 0 & 0                & 0
\end{pmatrix},\\[6pt]
J(b)&=\\*
&\hspace{-1.2cm}\begin{pmatrix}
 2 b-6 \Nc+\frac{6}{\Nc} & 0 & 0 & 0 & 0 & 2 \\[6pt]
 0                       & 2 b-6 \Nc+\frac{6}{\Nc} & 0                             & 0                     & 0                     & 2 \\[6pt]
 0                       & 0                       & 2 b+2 \Nc-\frac{2}{\Nc}       & 48                    & 48                    & 0 \\[6pt]
 0                       & 0                       & \frac{1}{2}-\frac{1}{2 \Nc^2} & 2 b+\frac{6}{\Nc}     & 0                     & \frac{2}{\Nc}-\frac{\Nc}{2} \\[6pt]
 0                       & 0                       & \frac{1}{2}-\frac{1}{2 \Nc^2} & 0                     & 2 b+\frac{6}{\Nc}     & \frac{2}{\Nc}-\frac{\Nc}{2} \\[6pt]
 12-\frac{12}{\Nc^2}     & 12-\frac{12}{\Nc^2}     & 0                             & \frac{48}{\Nc}-12 \Nc & \frac{48}{\Nc}-12 \Nc & 2 b-4 \Nc-\frac{2}{\Nc}
\end{pmatrix}.\nonumber
\end{align}
\endgroup
\end{subequations}
The 13 columns and rows correspond to the non-diagonal massless operator basis $\op_1^{(2)}$ to $\op_{13}^{(2)}$ in the order they are numbered in \eqs{eq:minBasis}.
The subblock $A(\Nf=0)$ has already been found for pure gauge theory~\cite{Husung:2019ytz}.
The explicit dependence on the argument $b$ is introduced to handle the occurrence of $\Nf$ due to the renormalisation of the coupling differently than the $\Nf$ arising from the number of flavours involved in constructing the operator.
This will become important for the mixed action case.

For non-vanishing quark masses we additionally get the mixing subblocks
\begin{subequations}\label{eq:mixingMassive1set}
\begingroup\allowdisplaybreaks
\begin{align}
(4\pi)^2\AD[\ch,\mm]{2}&=\left(\hspace{-5px}\begin{array}{ccccccccccc}
 12 \Nc & 0 & 8 \Nc-\frac{8}{\Nc} & 0 & 0 & 0 & 0 & 0 & 0 & 0 & 0 \\[6pt]
 \frac{79 \Nc}{30}+\frac{11}{30 \Nc} & 0 & \frac{149 \Nc}{60}-\frac{149}{60 \Nc} & 0 & 0 & 0 & 0 & 0 & 0 & 0 & 0 \\[6pt]
 \frac{37 \Nc}{30}+\frac{83}{30 \Nc} & 2 \Nf & \frac{197 \Nc}{60}-\frac{197}{60 \Nc} & 0 & 0 & 0 & 0 & 0 & 0 & 0 & 0 \\[6pt]
 0 & 0 & 16 & 0 & 0 & 0 & 0 & 0 & 0 & 0 & 0 \\[6pt]
 0 & 0 & -16 & 0 & 0 & 0 & 0 & 0 & 0 & 0 & 0 \\[6pt]
 0 & 0 & \frac{8}{\Nc}-8 \Nc & 0 & 0 & 0 & 0 & 0 & 0 & 0 & 0 \\[6pt]
 0 & 0 & 8 \Nc-\frac{8}{\Nc} & 0 & 0 & 0 & 0 & 0 & 0 & 0 & 0
\end{array}\hspace{-5px}\right),\\[6pt]
(4\pi)^2\AD[\suv,\mm]{2}&=\left(\hspace{-5px}\begin{array}{ccccccccccc}
 8 & 0 & 4 & 0 & 0 & 0 & 0 & 0 & -16 \Nc \Nf & 0 & 0 \\[6pt]
 8 & 0 & 4 & 0 & 0 & 0 & 0 & 0 & 0 & 0 & 0 \\[6pt]
 -32 & 0 & 48 & 0 & 0 & 0 & 0 & 0 & 0 & 0 & 0 \\[6pt]
 \frac{4}{\Nc} & 0 & \frac{2}{\Nc}-2 \Nc & 0 & 0 & 0 & 0 & 0 & 0 & 0 & 0 \\[6pt]
 \frac{4}{\Nc} & 0 & \frac{2}{\Nc}-2 \Nc & 0 & 0 & 0 & 0 & 0 & 0 & 0 & 0 \\[6pt]
 -\frac{16}{\Nc} & 0 & \frac{24}{\Nc}-24 \Nc & 0 & 32 \Nf & 0 & 0 & 0 & 0 & 0 & 0
\end{array}\hspace{-5px}\right),\\[6pt]
(4\pi)^2\AD[\mm]{2}&=\begin{pmatrix}
\mathfrak{M} & \NULL_{4\times 4} & \NULL_{4\times 3}\\[6pt]
\NULL_{4\times 4} & \mathfrak{M} & \NULL_{4\times 3}\\[6pt]
\NULL_{3\times 4} & \NULL_{3\times 4} & \left(6 \Nc-\frac{6}{\Nc}\right)\mathbb{1}_{3\times 3}
\end{pmatrix},\\[6pt]
\mathfrak{M}&=\begin{pmatrix}
 4 \Nc-\frac{8}{\Nc} & -2 \Nf & \frac{3}{\Nc}-3 \Nc & 0 \\[6pt]
 0 & -\frac{4 \Nc}{3}-\frac{6}{\Nc}+\frac{4 \Nf}{3} & 0 & 6 \Nc-\frac{6}{\Nc} \\[6pt]
 0 & 0 & 6 \Nc-\frac{6}{\Nc} & 0 \\[6pt]
 0 & 0 & 0 & 6 \Nc-\frac{6}{\Nc}
\end{pmatrix}.\nonumber
\end{align}
\endgroup
\end{subequations}
Once a specific choice for $\Nf>0$ and $\Nc$ is being made, the diagonalisation can be performed using the \texttt{Mathematica} notebook provided~\cite{HusungThesis}.
Also the spectrum of the 1-loop coefficients of the anomalous dimension matrix depends on these values and the symbolic expressions are very complicated if one does not fix $\Nf$ and $\Nc$ to some numerical values.
Therefore we refrain from trying to give here the symbolic results of a diagonal basis altogether and point again to said \texttt{Mathematica} notebook~\cite{HusungThesis}.

\subsection{Generalisation to actions with two flavour subsets}\label{sec:ADmixed}
Instead of considering just mixed actions, where one flavour subset is quenched, we will discuss the fully general case of two flavour subsets $q$ and $Q$ using different lattice fermion actions.

At mass-dimension~5, this doubles the massless operator basis without introducing any new mixing.
However, in the presence of massive operators the minimal basis is severely increased, but still the same three distinct eigenvalues from \eq{eq:spectrum[O]=5} must be considered for the spectrum.
We thus skip stating the entire mixing matrix at mass-dimension~5 as it does not yield much new insight.

At mass-dimension~6 however, having two distinct flavour subsets requires the inclusion of an additional kind of 4-fermion operators already for the massless case, like e.g.~for the mixed action in \eq{eq:mixedActionBasis}.
These operators give rise to new contributions in the spectrum.
Allowing for the absence of quenching, we then find for the enlarged massless operator basis in \eq{eq:mixedActionBasis} the nonzero subblocks of the mixing matrix\footnote{Using $\AD[\ch]{2}$ as orientation, the block rows (columns) are ordered from top (left) to bottom (right) as:
pure gauge operators, $\bar{q}\gamma_\mu D_\mu^3q$, $\bare{g}^2(\bar{q}\Gamma_i^\text{chiral}q)^2$, $\bar{Q}\gamma_\mu D_\mu^3Q$, $\bare{g}^2(\bar{Q}\Gamma_i^\text{chiral}Q)^2$, $\bare{g}^2(\bar{q}\Gamma_i^\text{chiral}q)(\bar{Q}\Gamma_i^\text{chiral}Q)$.
The ordering of the flavour subsets is the same in $\AD[\suv]{2}$, but there are only 4-fermion operators present.
Overall ordering of the various $\Gamma_i$ insertions in the 4-fermion operators remains unchanged compared to the single set of flavours.}
\begin{subequations}
\begingroup\allowdisplaybreaks
\begin{align}
(4\pi)^2\AD[\ch]{2}&=\begin{pmatrix}
 A(\Nf)            & B                & C                & B                & C                & 2C            \\[6pt]
 D(\Nf^q)          & E                & F(\Nf^q)         & 0                & K(\Nf^q)         & F(2\Nf^q)     \\[6pt]
 \NULL_{4\times 2} & \NULL_{4\times1} & G(\betaN,\Nf^q)  & \NULL_{4\times1} & \NULL_{4\times4} & L(\Nf^q)      \\[6pt]
 D(\Nf^Q)          & 0                & K(\Nf^Q)         & E                & F(\Nf^Q)        & F(2\Nf^Q)     \\[6pt]
 \NULL_{4\times2}  & \NULL_{4\times1} & \NULL_{4\times4} & \NULL_{4\times1} & G(\betaN,\Nf^Q) & L(\Nf^Q)      \\[6pt]
 \NULL_{4\times2}  & \NULL_{4\times1} & P(\Nf^Q)         & \NULL_{4\times1} & P(\Nf^q)        & R(\betaN,\Nf)
\end{pmatrix},\\[6pt]
(4\pi)^2\AD[\suv]{2}&=\begin{pmatrix}
J(\betaN)        & \NULL_{6\times6} & \NULL_{6\times6} \\[6pt]
\NULL_{6\times6} & J(\betaN)        & \NULL_{6\times6} \\[6pt]
\NULL_{6\times6} & \NULL_{6\times6} & J(\betaN)
\end{pmatrix},\\[6pt]
(4\pi)^2\AD[\suv,\ch]{2}&=\begin{pmatrix}
\NULL_{6\times3} & H                & \NULL_{6\times1} & \NULL_{6\times4} & H \\[6pt]
\NULL_{6\times3} & \NULL_{6\times4} & \NULL_{6\times1} & H                & H \\[6pt]
\NULL_{6\times3} & \NULL_{6\times4} & \NULL_{6\times1} & \NULL_{6\times4} & \NULL_{6\times4}
\end{pmatrix}.
\end{align}
\endgroup
\end{subequations}
Here $\Nf=\Nf^q+\Nf^Q$ and we introduced further short-hands
\begin{subequations}\label{eq:mixingShorthandsMixedAction}
\begingroup\allowdisplaybreaks
\begin{align}
K(\Nf)&=
\begin{pmatrix}
 0 & 0 & -\frac{7 \Nf}{30} & 0
\end{pmatrix},\\[6pt]
L(\Nf)&=
\begin{pmatrix}
 0 & 0 & -\frac{8}{3}                    & 0 \\[6pt]
 0 & 0 & -\frac{8}{3}                    & 0 \\[6pt]
 0 & 0 & \frac{8 \Nf}{3}-\frac{4}{3 \Nc} & 0 \\[6pt]
 0 & 0 & -\frac{4}{3 \Nc}                & 0
\end{pmatrix},\\[6pt]
P(\Nf)&=
\begin{pmatrix}
 0 & 0 & 0               & 0 \\[6pt]
 0 & 0 & 0               & 0 \\[6pt]
 0 & 0 & \frac{4 \Nf}{3} & 0 \\[6pt]
 0 & 0 & 0               & 0
\end{pmatrix},\\[6pt]
R(b,\Nf)&=
\begin{pmatrix}
 2 b               & 0                 & 0                         & -12                  \\[6pt]
 0                 & 2 b               & -12                       & 0                    \\[6pt]
 0                 & \frac{3}{\Nc^2}-3 & 2 b-3 \Nc+\frac{4 \Nf}{3} & 3 \Nc-\frac{12}{\Nc} \\[6pt]
 \frac{3}{\Nc^2}-3 & 0                 & 3 \Nc-\frac{12}{\Nc}      & 2 b-3 \Nc
\end{pmatrix}.
\end{align}
\endgroup
\end{subequations}
The extension to the massive operator basis gives rise to no new values in the spectrum apart from those already found for the non-mixed action case.
Of course some values in the spectrum now have a higher degeneracy or, if they are $\Nf^{q,Q}$ dependent, will split into two different values.
We omit stating the full massive mixing matrix to keep the results somewhat compact.
Keep in mind that this additional mixing is one-directional such that it does not affect the spectrum for $\hat{\gamma}_i$ already found in the massless case.
Also, the spectrum $\hat{\gamma}_i$ for the full massive case is known but some matching coefficients of the massive operators are unknown, i.e.~some tree-level matching coefficients may vanish thus suppressing those contributions by at least one power in $\gbar^2(1/a)$ further.

\subsection{Renormalisation of contact term interactions}
When considering unimproved Wilson fermions relying on maximal chiral twist to achieve automatic $\ord(a)$ improvement\footnote{The same argument actually holds for general massless Wilson quarks.} due to the continuum $T_1$-symmetry from \eq{eq:T1symmTransf}, also contact terms from double insertions of the allowed mass-dimension~5 operators $\mu^2\bar{\chi}\chi$ and $i/4\bar{\chi}\sigma_{\mu\nu}F_{\mu\nu}\chi$ come into play at $\ord(a^2)$.
Since we cannot access the massive mixing in the chirally twisted theory without repeating the entire renormalisation procedure in the twisted theory, we will restrict considerations to the renormalisation of the double insertion of $i/4\bar{\chi}\sigma_{\mu\nu}F_{\mu\nu}\chi$ in the massless case.
Working in the massless theory ensures that any chiral twist does not affect the continuum QCD action and we thus can reuse the setup of the untwisted theory.
For dimensional reasons, $i/4\bar{\chi}\sigma_{\mu\nu}F_{\mu\nu}\chi$ is anyway the only double insertion that can affect the matching coefficients of the massless mass-dimension~6 operator basis.
This contribution remains unchanged when going to the massive theory due to working in a mass-independent renormalisation scheme.

\begin{figure}
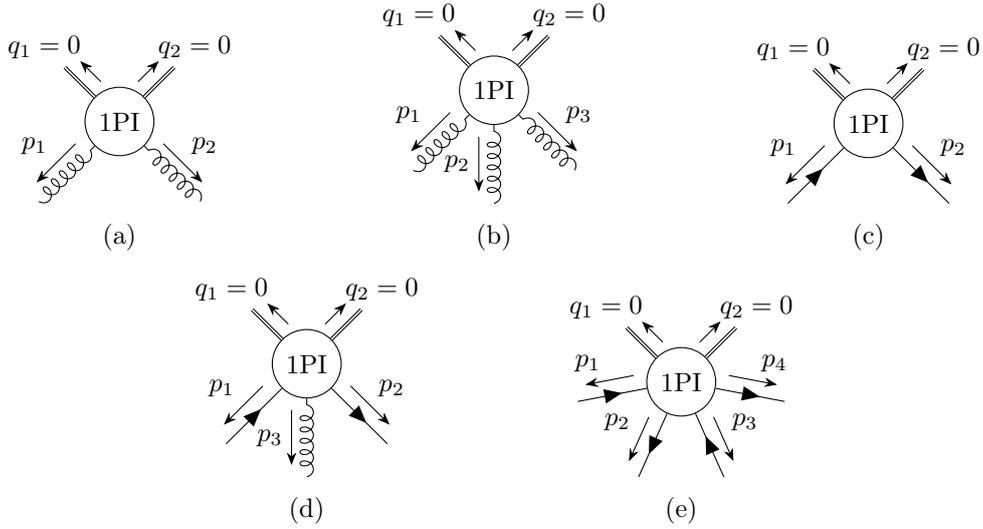

\centering
\begin{subfigure}[t]{0.32\textwidth}\centering
\includegraphics[]{\imgPath A2OO.pdf}
\caption{}
\label{fig:B2OO}
\end{subfigure}
\begin{subfigure}[t]{0.32\textwidth}\centering
\includegraphics[]{\imgPath A3OO.pdf}
\caption{}
\label{fig:B3OO}
\end{subfigure}
\begin{subfigure}[t]{0.32\textwidth}\centering
\includegraphics[]{\imgPath F2OO.pdf}
\caption{}
\label{fig:F2OO}
\end{subfigure}
\vskip5pt
\begin{subfigure}[t]{0.32\textwidth}\centering
\includegraphics[]{\imgPath F2AOO.pdf}
\caption{}
\label{fig:F2BOO}
\end{subfigure}
\begin{subfigure}[t]{0.32\textwidth}\centering
\includegraphics[]{\imgPath F4OO.pdf}
\caption{}
\label{fig:F4OO}
\end{subfigure}
\caption{Graphical representation of all considered 1PI graphs when renormalising the UV poles arising from contact interactions of double operator insertions.
Each operator is again inserted via an ``anchor'' field, which is indicated via the double line.
Both insertions are at zero momentum $q_1=q_2=0$.}\label{fig:contactRenorm}
\end{figure}

To determine what impact this double insertion has on the matching coefficients, we then need to renormalise the UV divergences arising from the contact interactions.
Consequently we also consider the 1PI graphs in \fig{fig:contactRenorm}, which gives the additional mixing contributions $\delta Z$
\begin{equation}
[\tilde{\op}_1^{(1)}(0)\tilde{\op}_1^{(1)}(0)]_{\MSbar}=Z_{1i}^{(1)}Z_{1j}^{(1)}\tilde{\op}_i^{(1)}(0)\tilde{\op}_j^{(1)}(0)+\delta Z_{k}\tilde{\op}_k^{(2)}(0).\label{eq:contactTermRenorm}
\end{equation}
The appropriate mass-dimension of the mixing operator $\op_k^{(2)}$ is fixed by the canonical mass-dimension of $\op_1^{(1)}$ since $\epsilon=(4-D)/2$ and thus $\delta Z$ is dimensionless, which does not allow mixing with operators of another mass-dimension.
Of course $\op_k^{(2)}$ would still contain massive operators of appropriate mass-dimension, if we considered the massive case.

Formally, the resulting additional mixing can be implemented by adding one row (and a trivial column, see \eq{eq:contactTermRenorm}) to the existing mixing matrix.
The double insertion will only get mixing contributions from the existing minimal operator basis and not introduce any new mixing beyond that.
We restrict considerations to the case, where only one flavour subset contributes a double insertion of $i/4\bar{Q}\sigma_{\mu\nu}F_{\mu\nu}Q$, e.g.~when using twisted Wilson fermions as valence quarks.

For the massless case and two flavour subsets, we then find the mixing contributions
\begin{align}
\delta Z &= \begin{pmatrix}
\NULL_{1\times 3} & \Delta(\Nf^Q) & 0 & \delta^{\ch}(\Nf^Q) & \delta^{\ch}(2\Nf^Q) & \NULL_{1\times6} & \delta^{\suv} & \NULL_{1\times6}
\end{pmatrix}\frac{\gbar^2}{(4\pi)^2\epsilon}\nonumber\\
&\hphantom{=}+\ord(\gbar^4),\label{eq:PsiSigmaFPsiContact}\\
\Delta(\Nf) &= \begin{pmatrix}
0 & 0 & -\frac{\Nf}{12} & 0
\end{pmatrix},\nonumber\\
\delta^{\ch}(\Nf) &= \begin{pmatrix}
0 & \frac{3}{8}-\frac{3}{8 \Nc^2} & \frac{7\Nc}{24}+\frac{1}{3 \Nc}-\frac{\Nf}{12} & \frac{3}{2 \Nc}-\frac{3 \Nc}{8}
\end{pmatrix},\nonumber\\
\delta^{\suv} &= \begin{pmatrix}
\frac{3}{2 \Nc^2}-\frac{3}{2} & 0 & 0 & \frac{3 \Nc}{2}-\frac{6}{\Nc} & 0 & \frac{\Nc}{16}
\end{pmatrix},\nonumber
\end{align}
where the columns correspond to the same massless operator basis used in \sect{sec:ADmixed}.
As we can see, 4-fermion operators both preserving and breaking chiral symmetry are required to renormalise the contact interactions, such that these operators will have non-vanishing tree-level matching coefficients in the basis in Jordan normal form as introduced in the following subsection.

\subsection{Renormalisation Group Invariants and the asymptotic lattice spacing dependence}\label{sec:RGIs}
For a generic Renormalisation Group invariant (RGI) spectral quantity $\obs$, the asymptotic lattice spacing dependence is, see also~\cite{Husung:2019ytz},
\begin{equation}
\obs(a)=\obs(0)-a^{\nmin}\sum_ic_i^\op(\gbar)\delta\obs_i^\op(1/a)+\ord(a^{\nmin+1}),
\end{equation}
where $\delta\obs_i^\op(1/a)$ is the contribution of the operator $\op_i$ to the lattice artifacts renormalised at scale $\mu=1/a$ and
\begin{equation}
c_i^\op(\gbar)=\omega_j^\op Z_{ji}^{-1}=\bar{c}_i^\op+\ord(\gbar^2)
\end{equation}
are the matching coefficients with tree-level value $\bar{c}_i^\op$.
The remaining lattice spacing dependence of $\delta\obs_i^\op(1/a)$ is dictated by the Renormalisation Group equation~(RGE)
\begin{equation}
\mu\frac{\rmd \delta\obs_i^\op(\mu)}{\rmd\mu}=\gamma_{ij}^\op(\gbar)\delta\obs_j^\op(\mu)
\end{equation}
and thus depends only on the anomalous dimensions of our minimal operator basis.
A formal solution to the RGE is
\begin{equation}
\delta\obs_i^\op(\mu)=\Pexp\left[\int\limits_{\gbar(\mu_0)}^{\gbar(\mu)}\rmd x\,\frac{\gamma^\op(x)}{\beta(x)}\right]_{ij}\delta\obs_j^\op(\mu_0),
\end{equation}
where $\beta(\gbar)=-\gbar^3(b_0+\ord(\gbar^2))$ is the $\beta$-function and $\Pexp$ denotes the path-ordered exponential with increasing $x$ from the right to the left.
The path-ordering is needed due to mixing of the operators under renormalisation, i.e.~$\gamma^\op$ is in general not diagonal.

Next we switch the basis $\op\rightarrow \base$ such that our 1-loop anomalous dimension matrix $\gamma_0^\base$ has Jordan normal form, which takes care of the fact that $\gamma_0^\op$ can be a non-diagonalisable matrix.
While for the cases of $\Nf>0$ or $\Nf^{q,Q}>0$ considered here this is not an issue and the Jordan normal form becomes just a diagonal matrix, both quenched and mixed actions can yield a non-diagonalisable 1-loop mixing matrix and thus require special care.
To pull out $\gamma_0^\base$ from the path ordered exponential we need to solve another RGE\footnote{We thank Agostino Patella for pointing out the issue with path-ordering at subleading orders.}\kern-0.5em, see~\cite{Papinutto:2016xpq},
\begin{equation}
\mu\frac{\rmd W(\mu)}{\mu}=\comm{\gamma^\base(\gbar)}{W(\mu)}-\beta(\gbar)\left\{\frac{\gamma^\base(\gbar)}{\beta(\gbar)}-\frac{\gamma_0^\base}{b_0\gbar(\mu)}\right\}W(\mu)
\end{equation}
with leading order solution $W(\mu)=\mathbb{1}+\ord(\gbar^2)$.
This allows to rewrite
\begin{align}
\delta\obs_i^\base(\mu)&=W^{-1}_{ij}(\mu)\exp\left[\frac{\gamma_0^\base}{b_0}\ln\left(\frac{\gbar(\mu)}{\gbar(\mu_0)}\right)\right]_{jk}W_{kl}(\mu_0)\delta\obs_l^\base(\mu_0),
\end{align}
where we used that $\gamma_0^\base$ is in Jordan normal form and thus can be written as a diagonal matrix plus one matrix containing only one off-diagonal.
Both matrices forming $\gamma_0^\base$ commute with each other, such that the path-ordering plays no role here.
Eventually we can introduce the Renormalisation Group Invariants
\begin{align}
\delta\RGI[i;]\obs^\base&=\lim_{\mu\rightarrow\infty}[2b_0\gbar^2(\mu)]_{ij}^{-\hat{\gamma}}\exp\left[\left\{\hat{\gamma}-\frac{\gamma_0^\base}{2b_0}\right\}\ln\left(2b_0\gbar^2(\mu)\right)\right]_{jk}\delta\obs_k^\base(\mu),
\end{align}
where we introduced
\begin{equation}
\hat{\gamma}=\diag\left(\frac{\gamma_0^\base}{2b_0}\right).
\end{equation}
The factor $2b_0$ in front of $\gbar^2$ is the common choice for the normalisation.
Finally this allows to rewrite
\begin{align}
\delta\obs_i^\base(\mu)&=W^{-1}_{ij}(\mu)[2b_0\gbar^2(\mu)]^{\hat{\gamma}}_{jk}\exp\left[\left\{\frac{\gamma_0^\base}{2b_0}-\hat{\gamma}\right\}\ln\left(2b_0\gbar^2(\mu)\right)\right]_{kl}\delta\RGI[l;]\obs^\base\label{eq:asympRGI}\\
&=[2b_0\gbar^2(\mu)]^{\hat{\gamma}}_{ij}\exp\left[\left\{\frac{\gamma_0^\base}{2b_0}-\hat{\gamma}\right\}\ln\left(2b_0\gbar^2(\mu)\right)\right]_{jk}\delta\RGI[k;]\obs^\base\times\left(1+\ord(\gbar^2)\right),\label{eq:asympRGIexpanded}
\end{align}
where now all scale dependence is absorbed into the prefactor of the RGI quantity.
By construction $\delta\RGI[k;]\obs^\base$ is scale-independent.

In the cases of $\Nf>0$ or $\Nf^{q,Q}>0$ considered here the 1-loop anomalous dimension matrix can be diagonalised such that the remaining exponential in \eqs{eq:asympRGI} and \eqref{eq:asympRGIexpanded} becomes the identity.
For both quenched and mixed actions we find terms with additional factors of $\ln(2b_0\gbar^2)$ modifying the simple power law $[\gbar^2(1/a)]^{\hat{\gamma}}$.
These logs will in general give the dominating contributions for operators with the given leading power $\hat{\gamma}$ in the coupling.
However, these particular contributions belong here to chiral symmetry violating 4-fermion operators that are typically suppressed by one power in the coupling as will be discussed in the following section.

\def\spectraPlots{\figs{fig:spectraWilsonGW} to \ref{fig:spectraMixed}}

\section{Matching to lattice actions}
\label{sec:matching}
With the full 1-loop mixing matrix computed in the previous section we are now able to compute the leading powers in the coupling modifying the naive $a^n$ behaviour.
While in general these leading powers are greater or equal to $\hat{\gamma}_i^{(n)}=(\gamma^\base_0)^{(n)}_{ii}/(2b_0)$, obtained from the 1-loop coefficients of the mixing matrices listed in section~\ref{sec:AD}, we actually may gain additional insight from (tree-level) matching.
This means we want to determine the tree-level values of the matching coefficients $\bar{c}_i^\op\equiv \omega_i^\op(0)$ in \eq{eq:matchingCoeffs} or more importantly its counterpart for the diagonal basis~$\bar{c}_i^\base$.
If a tree-level matching coefficient vanishes for an operator of our diagonalised basis, it shifts the associated leading power in the coupling by (at least) one power in the coupling further.
Tree-level matching has the particularly beneficiary feature that a naive expansion of the lattice action in the lattice spacing suffices to obtain the desired tree-level matching coefficients.
Beyond tree-level, the perturbative matching procedure requires $l$-loop computations in perturbation theory for both the lattice theory and continuum SymEFT to achieve matching to $l$-loops.
Especially the lattice side then becomes very complicated due to relying on lattice perturbation theory.

We therefore take only tree-level matching into account, when computing the spectrum of leading powers in the coupling.
From tree-level matching we can only infer, whether a leading power in the coupling vanishes and if so assume the next-to-leading order (NLO) to be the truly leading power in the coupling.
We then introduce
\begin{equation}
\hat{\Gamma}_i=\hat{\gamma}_i+n_i
\end{equation}
as the actual leading power in $\bar{g}^2(1/a)$, where $n_i$ takes $(n_i-1)$-loop improvement into account, i.e.~$n_i=1$ implies absence of contributions from this operator at tree-level.
The various powers of $\gbar^2(1/a)$ for the different operators contributing to the lattice artifacts may spread over more than one power in the coupling, such that e.g.~the subleading contributions of one operator compete with the leading power contributions of another operator.

To match the coefficients of the 4-fermion operators at tree-level, we must first understand what tree-level in terms of our 4-fermion operators actually means.
For this it pays out to draw possible connected fermion 4-point functions at tree-level using (some) lattice Feynman rules, in which case the leading order will always be $\ord(g^2)$ just like our particular choice of prefactor included in the definition of our 4-fermion operator basis in \eqs{eq:4fermionChiral} and \eqref{eq:4fermionNonChiral}.
Since there are no UV divergences present at tree-level and thus no renormalisation needed to this order in perturbation theory, we immediately know that we could equally well consider tree-level vertex functions to perform the matching.
This shows that all tree-level matching coefficients of 4-fermion operators in the basis $\opSix_i$ are zero as there is no 4-fermion vertex in any of the considered lattice actions.
There is however one caveat as a general lattice gauge action might give rise to a term $\frac{1}{\bare{g}^2}\tr([D_{\mu},F_{\mu\rho}][D_{\nu},F_{\nu\rho}])$, which then by EOMs will be absorbed into the 4-fermion operator~$\opSix_7$.

\subsection{Tree-level matching for $\ord(a)$ improved massless action}
We find as tree-level matching coefficients for the massless mass-dimension~6 operator basis $\opSix_i$ for a fully on-shell $\ord(a)$ improved lattice action
\begin{equation}
\bar{c}^\op_1=\frac{e_2}{3},\quad \bar{c}^\op_2=e_1-e_3+\frac{1}{12},\quad \bar{c}^\op_3=\frac{1}{6},\quad \bar{c}^\op_{4-5}=0,\quad \bar{c}^\op_6=-\frac{e_2}{3}-e_3,\quad \bar{c}^\op_{7-13}=0,
\end{equation}
where $\bar{c}^\op_3$ is universal for the action of either Wilson, Overlap or Domain wall quarks as they eventually all employ the same Wilson Dirac operator.
Both tree-level coefficients $\bar{c}^\op_1$ and $\bar{c}^\op_2$ for the gluonic operators have been taken from~\cite{Luscher:1984xn}, where $e_i$ are the coefficients for different terms in the lattice action\footnote{We renamed those coefficients to $e_i$ to avoid a clash with our notation.}\kern-0.5em~with the conventional normalisation $e_0+8e_1+8e_2+16e_3=1$, namely the plaquette~($e_0$), rectangle~($e_1$), chair~($e_2$) as well as twisted chair~($e_3$) Wilson loops.
This general class of pure gauge actions covers a wide range of possible lattice gauge actions.
A common and natural choice is to set $e_2=e_3=0$, which reduces the computational effort as only two kinds of Wilson loops must be computed.
It also sets $\bar{c}^\op_6=0$, which is interesting as now only two coefficients remain nonzero, namely
\begin{equation}
\bar{c}^\op_2=e_1+\frac{1}{12},\quad \bar{c}^\op_3=\frac{1}{6}, \quad \bar{c}^\op_{1,4-13}=0.\label{eq:commonMatching}
\end{equation}
A particularly useful choice for the remaining coefficient of the Wilson loops in the lattice gauge action is $e_1=-1/12$, which is known as the L\"uscher-Weisz action~\cite{Luscher:1984xn}.
It ensures tree-level $\ord(a^2)$ improvement of the lattice gauge action such that only $\bar{c}^\op_3$ remains nonzero.
We already discussed this for lattice pure gauge theory~\cite{Husung:2019ytz}.

Having a vanishing tree-level coefficient in the basis $\opSix_i$ only ensures overall absence, if the corresponding operator does not mix into another operator that has a non-vanishing tree-level coefficient or mixes itself into another operator and so on as the (tree-level) matching coefficients of the basis in Jordan normal form can be obtained through
\begin{equation}
c_i^\base=c_j^\op T_{ji}^{-1},\quad \baseSix_i=T_{ij}\opSix_j,\label{eq:diagonalMatching}
\end{equation}
where $T$ is the change of basis such that the 1-loop mixing matrix is in Jordan normal form, or diagonal as in most cases discussed here.
All chiral-symmetry breaking 4-fermion operators have vanishing tree-level matching coefficients.

The chiral-symmetry preserving 4-fermion operators can mix into any other massless operator present in our on-shell basis, such that they will generally be present at tree-level, if any other tree-level matching coefficients are nonzero in the basis $\opSix_i$.

Knowing the TL matching coefficients has already very interesting consequences.
Assuming use of only the plaquette and rectangle term in the gauge action, i.e.~the matching coefficients from \eq{eq:commonMatching}, we will find in the massless case only the two operators $\opSix_{2}$ and $\opSix_3$ to have non-vanishing tree-level matching coefficients, when considering non-perturbatively $\ord(a)$ improved Wilson quarks as well as Ginsparg-Wilson quarks.
In particular the choice of the tree-level improved L\"uscher-Weisz lattice gauge action~\cite{Luscher:1984xn} will ensure that only $\opSix_3$ has then a non-vanishing tree-level matching coefficient.
This allows either to parametrise the leading lattice artifacts in terms of this single coefficient or alternatively to perform Symanzik tree-level improvement at $\ord(a^2)$ by adding only one additional term to the lattice fermion action like e.g.
\begin{equation}
\delta S_\mathrm{latt}=-\frac{a^6}{12}\sum_{\mu,x}\bar{\Psi}(x)\gamma_\mu\left\{\nabla_\mu+\nabla_\mu^*\right\}\nabla_\mu^*\nabla_\mu\Psi(x),
\end{equation}
see also \cite{Alford:1996pk,Alford:1996nx}, where this has already been pointed out.

For a more in depth discussion of the (massive) tree-level matching coefficients and their impact on the leading lattice artifacts for $\ord(a)$ improved Wilson quarks, Ginsparg-Wilson quarks and Domain-Wall fermions, we refer the reader to~\cite{Husung:2021mfl}.

\subsection{Tree-level matching for chirally twisted Wilson quarks}
In case there are double insertions of mass-dimension~5 operators present that break chiral symmetry, as is the case for twisted mass QCD relying on automatic $\ord(a)$ improvement, additional $\ord(a^2)$ corrections occur.
These double insertions then give rise to contact terms in the SymEFT, whose renormalisation affect the tree-level matching coefficients of the 4-fermion operators, including chiral-symmetry violating ones, in the basis $\baseSix_i$ as we can see in \eq{eq:PsiSigmaFPsiContact}.

Taking this generalisation into account is achieved in a very sloppy way by enlarging the mixing matrix and thus $T$ in \eq{eq:diagonalMatching} by one row and column (in the massless case) formed by \eq{eq:PsiSigmaFPsiContact}.
Apart from this, the general remarks from the $\ord(a)$ improved case remain the same.
Regarding the matching for the massive twisted theory, we lack some information as currently only the massless mixing matrix is known.
This allows to determine $\hat{\gamma}_i^{(2)}$ for all operators including the massive ones, but we do not know, whether any tree-level matching coefficients of the massive operators in the basis $\baseSix_i$ vanish.
We thus assume they do not vanish.

\subsection{Leading powers in the coupling incorporating tree-level matching}
Taking all of this into account, we find the leading powers and up to next-to-next-to-next-to-leading orders (N${}^3$LO) in the coupling, here denoted as $\hat{\Gamma}_i$, for the various cases as depicted in \spectraPlots.
Going that far into the subleading powers is done only to highlight the overall spread of the leading powers from the various contributions as some of them are severely suppressed by up to three powers in $\gbar^2(1/a)$ compared to the lowest power $\hat{\Gamma}_\mathrm{min}$, depending on $\Nf$.
There also the extension to the massive case is included, where of course the same considerations for tree-level matching can and should be applied to ensure sorting out any vanishing tree-level coefficients.
As we can see, the various powers in $\gbar^2(1/a)$ form a dense spectrum.
For use in an ansatz for the continuum extrapolation we give in \tab{tab:leadingPowers} the numerical values of the three (five) leading powers in $\gbar^2(1/a)$ for the massless (massive) case for the lattice discretisations discussed in this paper at $\Nc=3$ and various values of $\Nf$.
For other choices of $\Nf$ or $\Nc$ we refer the reader again to the \texttt{Mathematica} notebook~\cite{HusungThesis}.

\begin{table}
\caption{Non-exhaustive examples for leading powers in the coupling relevant for the leading order lattice artifacts $a^2[\gbar^2(1/a)]^{\hat\Gamma}$ at $\Nc=3$ and various values for $\Nf$.
Degeneracies are not listed as they cannot be distinguished numerically.
While vanishing matching coefficients have been taken into account by shifting the corresponding leading power by 1, hierarchies between the various coefficients are ignored and certainly should be looked at~\cite{Husung:2021mfl}.
We limit considerations to commonly used combinations and some interesting cases.
For the gauge action we assume use of a variant compatible to \eq{eq:commonMatching}.
The underlined numbers belong to massive operators.
If the L\"uscher-Weisz action~\cite{Luscher:1984xn} is being used rather than the more general case in \eq{eq:commonMatching} the leading powers do not change, since the operator affected by this change of matching is already sufficiently suppressed by $\hat{\gamma}_i\gtrsim 0.86$ for $2\leq\Nf\leq 8$.
The leading powers for $\ord(a)$ improved tmQCD should be identical to the Wilson case, but massive tree-level matching coefficients may differ and are currently unavailable.
We therefore assume those coefficients to be non-vanishing.}\label{tab:leadingPowers}
\begin{tabular}{lcl}
lattice discretisation                       & flavours & leading powers $\hat{\Gamma}$ \\\hline\hline
Ginsparg-Wilson                              & 0        & $0.273, \underline{0.424}, 0.597, 0.634, \underline{0.727}, \ldots$ \\
                                             & 2        & $\underline{-0.172}, 0.271, \underline{0.483}, 0.638, 0.711, \ldots$ \\
                                             & 3        & $\underline{-0.111}, 0.247, \underline{0.519}, 0.668, 0.760, \ldots$ \\
                                             & 4        & $\underline{-0.040}, 0.209, \underline{0.560}, 0.698, 0.817, \ldots$ \\\hline
$\ord(a)$ improved Wilson~\cite{Luscher:1996ug}                    & 0        & $0.273, \underline{0.424}, 0.597, 0.634, \underline{0.727}, \ldots$ \\
                                             & 2        & $\underline{-0.172}, 0.271, \underline{0.483}, 0.638, 0.711, \ldots$ \\
                                             & 3        & $\underline{-0.111}, 0.247, \underline{0.519}, 0.668, 0.760, \ldots$ \\
                                             & 4        & $\underline{-0.040}, 0.209, \underline{0.560}, 0.698, 0.699, \ldots$ \\\hline
mixed action, see e.g.~\cite{Bussone:2019mlt}& 2,2      & $\underline{-0.172}, 0.230, 0.271, \underline{0.483}, 0.638, \ldots$ \\
(Wilson sea quarks, tmQCD                    & 3,3      & $\underline{-0.111}, 0.198, 0.247, \underline{0.519}, 0.583, \ldots$ \\
valence quarks with SW term)                 & 4,4      & $\underline{-0.040}, 0.155, 0.209, \underline{0.560}, 0.580, \ldots$ \\\hline
$\ord(a)$ improved action                    & 2+1      & $\underline{-0.111}, 0.198, 0.247, \underline{0.519}, 0.583, \ldots$ \\
(light GW + heavy Wilson quarks)             & 3+1 or 2+2     & $\underline{-0.040},  0.155, 0.209, \underline{0.560}, 0.580,\ldots$ 
\end{tabular}
\end{table}

Considering e.g.~the case of $\Nf=3$ for $\ord(a)$ improved Wilson or GW quarks, even the lowest powers are only separated by roughly $\Delta\hat{\Gamma}\sim 0.4$ in the massless case.
For increasing number of flavours the truly leading power in the coupling then becomes clearly separated from the subleading powers, but at $\Nf=8$ it is negative.
For non-perturbatively $\ord(a)$ improved Wilson quarks the spectrum becomes even denser at subleading powers in the coupling due to the presence of (1-loop matching contributions of) chiral-symmetry breaking 4-fermion operators, but the leading power in the coupling remains the same for $\Nf\leq 8$ although the first subleading power may be less suppressed depending on the value of $\Nf$, see also \fig{fig:spectraWilsonGW}.

An interesting, pedagogical example are Wilson quarks at maximal chiral twist either relying on automatic $\ord(a)$ improvement at this twist angle or making explicit use of non-perturbative $\ord(a)$ improvement just like for non-twisted Wilson quarks.
Removing explicitly the $\ord(a)$ lattice artifacts eliminates the double operator insertions contributing to $\ord(a^2)$.
Thus no contact terms may arise ensuring that no chiral-symmetry violating 4-fermion operators are allowed at tree-level matching, which otherwise results in a slightly negative 1-loop anomalous dimension at $\Nf=2$, namely $\hat{\Gamma}_\mathrm{min}\approx-0.12$, becoming worse towards larger $\Nf$.
To compare both choices of $\ord(a)$ improvement, one can compare the spectrum in \fig{fig:spectratmQCD} with the case of Wilson in \fig{fig:spectraWilsonGW} at even number of flavours.
Evidently, the spectrum of leading powers in the coupling is denser without having explicit improvement.

As was probably to be expected, introducing two sets of flavours, either dynamical or quenched, introduces even more operators and thus severely increases the density of the spectrum compared to having only one lattice discretisation, see e.g.~\figs{fig:spectraWilsonGW} and \ref{fig:spectraMixed}.
For the cases considered here, this will also impact the leading power in the coupling for the massless case.
While e.g.~at $\Nf=3$ we only find a slight decrease from $\hat{\Gamma}_\mathrm{min}\approx 0.247$ to $\hat{\Gamma}_\mathrm{min}\approx 0.198$, this also abandons the already weakly pronounced gap of the 1st to 2nd leading power from $\Delta \hat{\Gamma}\approx0.42$ to $\Delta \hat{\Gamma}\approx0.05$.
This effect holds true for the other values for $\Nf$ considered here.
Notably, the case of a mixed action with two valence quarks considered here yields a non-diagonalisable mixing matrix at 1-loop order.
This gives rise to an additional factor $\log(2b_0\gbar^2(1/a))$ modifying the pure power law in the coupling for $\hat{\Gamma}\approx1.586$, which is however very suppressed compared to the leading powers in $\gbar^2(1/a)$, namely $\hat{\Gamma}_\mathrm{min}\approx 0.230$ in the massless case and $\hat{\Gamma}_\mathrm{min}\approx -0.172$ in the massive case.
The spectra of mixed action and light GW + heavy Wilson quarks in \fig{fig:spectraMixed} are very similar, but a few important differences occur for the latter case.
Namely the use of Fierz identities for $\Nf=2+1$ or $\Nf=3+1$ reduces the number of 4-fermion operators and having light GW quarks also excludes some chiral-symmetry violating 4-fermion operators as explained after \eq{eq:mixedActionBasis}.

In general, the additional contributions to the spectrum for the massive case result again in a slightly denser spectrum, but do not affect the spectrum of the massless operator basis due to the tridiagonal form of the mixing matrix, see \eq{eq:mixingMatrixStructure}.
In all cases considered, except twisted Wilson quarks purely relying on automatic $\ord(a)$ improvement, the lowest power found at $\Nf=2,3,4$ is always from the massive operator basis and actually slightly negative.
At $\Nf=8$ the lowest power is from the massless operator basis and negative as well.
Without counting double operator insertions, there are eleven distinct massive operators at mass-dimension~6, but only three distinct powers $\hat{\Gamma}_i$.
Taking the subleading powers into account, the spectrum of the massive operators modulo 1 has only two distinct powers.
This is due to a severe degeneracy of the spectrum for these operators since additional powers of the mass only shift the overall 1-loop anomalous dimension by a constant.
Introducing two sets of flavours does not affect the powers for the massive contributions as they do not depend on $\Nf^{q,Q}$ but $\Nf=\Nf^q+\Nf^Q$.
Of course the degeneracy in the spectrum grows even further.

From a numerical point of view this degeneracy suggests to treat the different operators contributing with the same power as one linear combination, because one cannot distinguish them during a fit anyway (actually most of the degeneracy in the power spectrum starts at subleading orders due to vanishing tree-level matching coefficients of some of the operators).
For tmQCD relying only on automatic $\ord(a)$ improvement, the spectrum for the massive case has two additional values, where one of those again increases the degeneracy.
Introducing the Wilson clover term with a non-perturbative improvement coefficient eliminates all truly new powers in the coupling and avoids any contact term renormalisation of our operator basis, which otherwise would affect $\ord(a^2)$ just like in the untwisted theory.
In particular, this ensures vanishing of tree-level matching coefficients for the chiral symmetry violating 4-fermion operators.
Only the double insertion of one massive operator remains, that has a degenerate power in the coupling and is thus indistinguishable from the contributions of massive mass-dimension~6 operators.

\def\scale{1}
\def\lastpage{6}
\begin{figure}
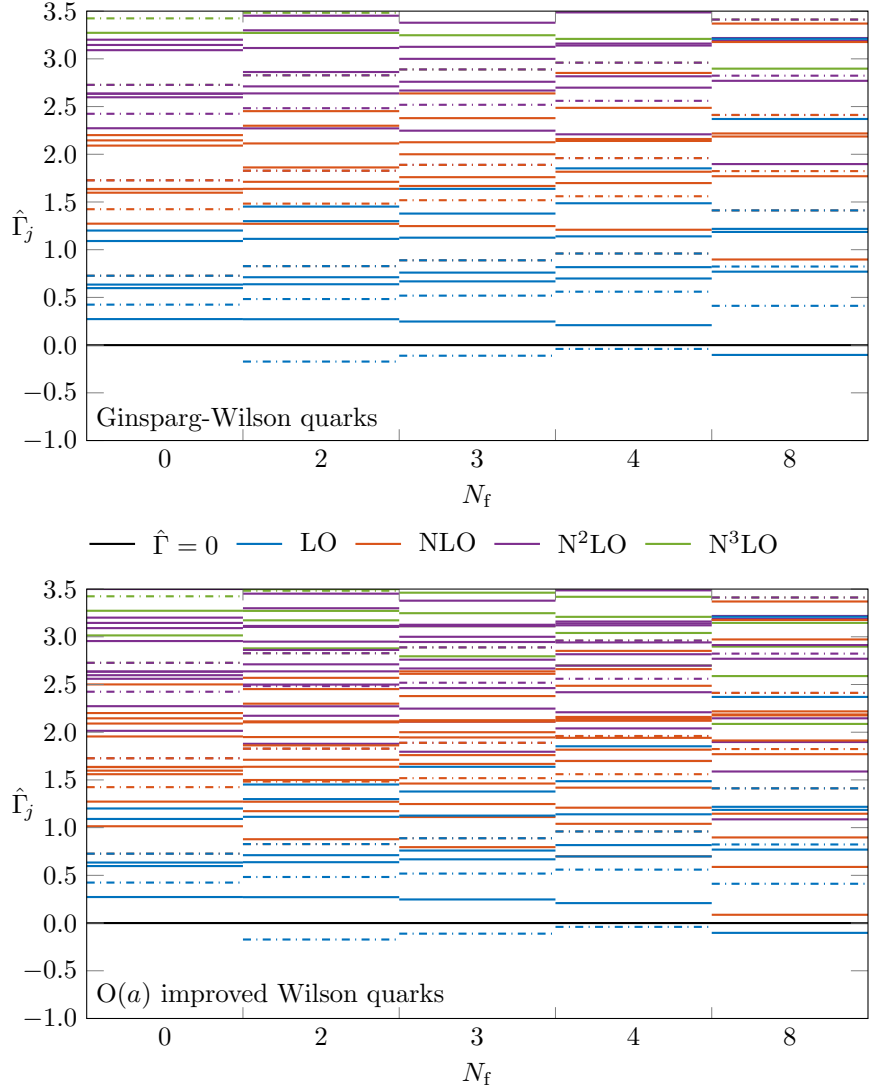
\centering
\includegraphics[scale=\scale,page=1]{\imgPath MassDimension6heatmaps.pdf}
\includegraphics[scale=1,page=\lastpage]{\imgPath MassDimension6heatmaps.pdf}
\includegraphics[scale=\scale,page=2]{\imgPath MassDimension6heatmaps.pdf}
\caption{Leading and subleading powers in the coupling $\hat{\Gamma}_j$ for the minimal on-shell operator basis of GW and Wilson quarks, describing the leading order lattice artifacts for spectral quantities at $\Nf=0,2,3,4,8$.
$\Nf=8$ is added to highlight what happens when one gets closer to the conformal window.
While the solid lines correspond to the massless operator basis (possibly containing massive mixing contributions), the dash-dotted lines belong to the operators with overall mass-dependence.
In case the tree-level coefficient vanishes, the first power plotted is regarded as NLO in the colour coding.
Due to overlapping numerical values leading powers may be hidden by the ``subleading'' powers of other operators.}
\label{fig:spectraWilsonGW}
\end{figure}

\begin{figure}
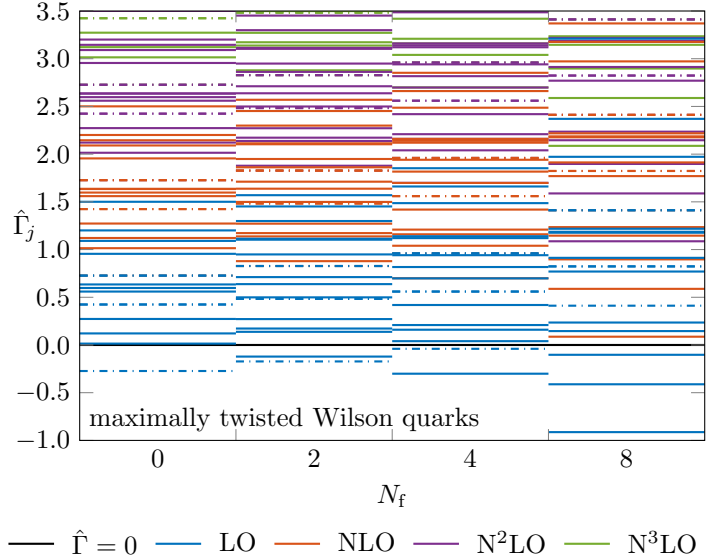
\centering
\includegraphics[scale=\scale,page=3]{\imgPath MassDimension6heatmaps.pdf}
\includegraphics[scale=1,page=\lastpage]{\imgPath MassDimension6heatmaps.pdf}
\caption{Leading and subleading powers in the coupling $\hat{\Gamma}_j$ for the minimal on-shell operator basis of Wilson quarks with maximal chiral twist and mass-degenerate flavour doublets.
Only the case without explicit $\ord(a)$ improvement is plotted as the case with improvement is identical to $\ord(a)$ improved Wilson quarks, see \fig{fig:spectraWilsonGW}.
These powers describe the leading order lattice artifacts for spectral quantities at $\Nf=0,2,4,8$.
$\Nf=8$ is added to highlight what happens when one gets closer to the conformal window.
Again the solid lines correspond to the overall massless operator basis (possibly containing massive mixing contributions), the dash-dotted lines belong to the operators with overall mass-dependence.
In case the tree-level coefficient vanishes the first power plotted is  regarded as NLO in the colour coding.
Due to overlapping numerical values leading powers may be hidden by the ``subleading'' powers of other operators.
Notice that we do not know the tree-level matching coefficients for the massive case, which are therefore assumed to be non-vanishing.
For the fully quenched case $\Nf=0$, the lowest power may be estimated too low as it could potentially be shifted to the first massless contribution instead.}
\label{fig:spectratmQCD}
\end{figure}

\begin{figure}
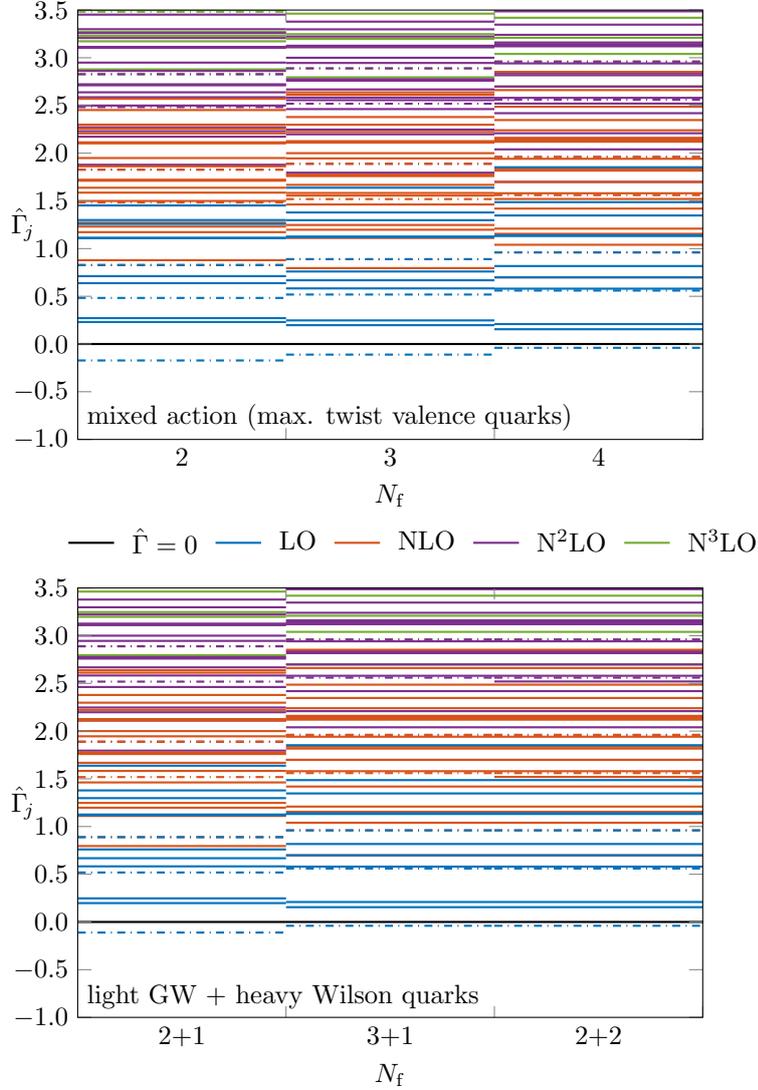
\centering
\includegraphics[scale=\scale,page=4]{\imgPath MassDimension6heatmaps.pdf}
\includegraphics[scale=1,page=\lastpage]{\imgPath MassDimension6heatmaps.pdf}
\includegraphics[scale=\scale,page=5]{\imgPath MassDimension6heatmaps.pdf}
\caption{Leading and subleading powers in the coupling $\hat{\Gamma}_j$ for the minimal on-shell operator basis describing the leading order lattice artifacts for spectral quantities at $\Nf=2,3,4$.
Here two distinct sets of flavours are assumed either in a mixed action with the setup described in~\cite{Bussone:2019mlt} or using different lattice descretisations for light and heavy quarks, here GW and Wilson quarks respectively.
While the solid lines correspond to the massless operator basis (possibly containing massive mixing contributions), the dash-dotted lines belong to the operators with overall mass-dependence.
In case the tree-level coefficient vanishes, the first power plotted is regarded as NLO in the colour coding.
Due to overlapping numerical values leading powers may be hidden by the ``subleading'' powers of other operators.}
\label{fig:spectraMixed}
\end{figure}

\section{Discussion}
We have computed the 1-loop anomalous dimensions of all mass-dimension~5 and 6 operators relevant for the minimal basis describing lattice artifacts from either the Wilson or Ginsparg-Wilson action up to and including $\ord(a^2)$.
This also includes all relevant massive operators as well as the necessary generalisation to the quenched case, mixed actions or use of two distinct lattice discretisation for different quark flavours.
These 1-loop coefficients of the anomalous dimension matrix modify the leading asymptotic lattice spacing dependence from the classical integer power $a^n$ behaviour to $a^n[\gbar^2(1/a)]^{\hat{\Gamma}_i^{(n)}}$.
Only for the quenched case and mixed actions considered here additional factors of $\log(\gbar^2(1/a))$ occur, see also section~\ref{sec:RGIs}.
Here $\hat{\Gamma}_i^{(n)}=\hat{\gamma}_i^{(n)}+n_i$ takes suppressions from matching coefficients by $\gbar^{2n_i}(1/a)$ into account.
In practice we only know whether the tree-level matching coefficients vanish for the lattice actions considered here except for the massive operators of tmQCD, where the full mixing with massive operators is currently not available.
We explicitly write here the additional superscript in $\hat{\Gamma}_i^{(n)}$, which has been dropped until now, to include also $\ord(a)$ corrections, i.e.~$n=1$.
We have discussed those effects already before~\cite{Husung:2019ytz} without taking the massive operators into account.
We only mention those effects briefly.

For example at $\Nf=3$ we found the lowest values of the massless spectrum to be $\hat{\Gamma}_\mathrm{min}^{(1)}\approx 0.074$ for the $\ord(a)$ lattice artifacts of unimproved Wilson quarks and $\hat{\Gamma}_\mathrm{min}^{(2)}\approx 0.25$ for the $\ord(a^2)$ lattice artifacts of both non-perturbatively $\ord(a)$ improved Wilson quarks and Ginsparg-Wilson quarks.
For typical numbers of flavours $\Nf=0,2,3,4$ all powers in the coupling from the massless operator basis improve the convergence towards the continuum limit as $a\searrow 0$ due to being positive.
Only maximally twisted Wilson quarks without explicit $\ord(a)$ improvement have e.g.~at $\Nf=2$ one slightly negative power in the coupling $\hat{\Gamma}_\mathrm{min}^{(2)}\approx -0.122$ for the massless operator basis arising from the contribution of 4-fermion operators that violate chiral symmetry.
Those contributions become worse for increasing number of flavours.
In contrast to the non-perturbatively $\ord(a)$ improved lattice actions those operators can arise here already at tree-level matching due to the renormalisation of contact terms from double insertions of the mass-dimension~5 operator basis in the SymEFT.

The example of twisted Wilson quarks with automatic $\ord(a)$ improvement actually teaches us another lesson, namely that relying on continuum symmetries for $\ord(a)$ improvement may worsen the approach to the continuum limit at $\ord(a^2)$ and beyond compared to explicit Symanzik improvement.
Performing instead explicit $\ord(a)$ improvement -- already tree-level improvement would suffice for that matter -- ensures that double operator insertions of mass-dimension~5 operators do not occur or at least occur suppressed by at least two additional powers in $\gbar^2(1/a)$, which then also shifts any effects from contact term renormalisation by this additional power in the coupling.
Fortunately the ETMC collaboration already performs explicit $\ord(a)$ improvement~\cite{Frezzotti:2005gi}.
Under this impression the earlier attempts of obtaining a classically perfect action, see e.g.~\cite{DeGrand:1995ji}, may have been a worthwhile endeavour, since this argument of course occurs at any order in the lattice spacing.
If there are negative powers in the coupling present at subleading powers in the lattice spacing, these will automatically be shifted up by at least one power in the coupling as will all $\hat{\Gamma}_i^{(n)}$ from tree-level contributions.

The additional massive operators must be considered, when one is working in a mass-independent renormalisation scheme rather than a hadronic scheme.
Overall they increase the density of the spectrum found for $\hat{\Gamma}^{(2)}_i$ even further, but introduce only three distinct powers due to a fairly degenerate spectrum.
In all cases considered here apart from twisted-mass QCD relying on automatic $\ord(a)$ improvement, the massive operators decrease the leading power in the coupling at $\Nf=2,3,4$ slightly compared to the massless basis.

In either case all the leading powers encountered are much better behaving than the dominant power $\min_i(\hat{\Gamma}_i^{(2)})=-3$ found in the 2-d O(3) sigma model~\cite{Balog:2009np,Balog:2009yj}, which is good news.
However, due to the large number of operators we get many different contributions $\hat{\Gamma}_i^{(2)}$ to the spectrum, which in addition lie in close proximity to one another, see also \spectraPlots.
This will make it very difficult to decide, which contributions actually dominate and therefore must be included in a proper fit-ansatz for continuum extrapolations.
The density in the spectrum for $\hat{\Gamma}_i^{(2)}$ becomes even worse when introducing two distinct lattice discretisations, like we discussed here in form of mixed actions or different discretisations for dynamical heavy and light quark flavours.
Eventually the situation is even more complicated since the sensitivity to contributions from different operators of our minimal basis will presumably depend on the (spectral) quantity to be extrapolated and also different orders of magnitude of the various LO matching coefficients will have an impact~\cite{Husung:2021mfl}.
Also, there are of course always corrections subleading in the power of the lattice spacing that one must be wary about.
All of this indicates that one should be very careful when doing continuum extrapolations regarding the systematic errors associated to the extracted continuum values.

\begin{table}\centering
\caption{Non-exhaustive overview of typical lattice spacings in today's lattice QCD simulations combined with the $\MSbar$ coupling at scale $\mu=1/a$ obtained from 5-loop running~\cite{Luthe:2017ttc}.
The MILC HISQ ensembles are only added for comparison as they involve staggered quarks, which were not included in our discussion.}\label{tab:typicalLatticeSpacings}
\begin{tabular}{l|c|c|c||c}
gauge action & Wilson & Iwasaki & L\"uscher-Weisz &  L\"uscher-Weisz\\
quark action & & Domain-Wall & $\ord(a)$ impr. Wilson &  MILC HISQ\\
$\Nf$ & $0$\cite{Husung:2017qjz} & $2+1$\cite{Boyle:2017jwu} & $2+1$\cite{Bruno:2017gxd} & $2+1+1$\cite{McLean:2019qcx}\\\hline\hline&&&\\[-9pt]
$a$ [fm] & $\begin{array}{cc}0.01 & 0.09\end{array}$ & $\begin{array}{cc}0.07 & 0.11\end{array}$ & $\begin{array}{cc}0.04 & 0.09\end{array}$ & $\begin{array}{cc}0.04 & 0.09\end{array}$\\[4pt]
$\alpha_{\overline{\text{MS}}}^{\text{5-loop}}(1/a)$ & $\begin{array}{cc}0.11 & 0.21\end{array}$ & $\begin{array}{cc}0.25 & 0.32\end{array}$ & $\begin{array}{cc}0.21 & 0.28\end{array}$ & $\begin{array}{cc}0.22 & 0.29\end{array}$
\end{tabular}
\end{table}

Before discussing possible extensions of this work we should ask ourselves whether today's lattice simulations are at sufficiently small lattice spacings such that a perturbative description of the leading asymptotic lattice spacing dependence suffices.
While there will never be absolute certainty that the leading power in the lattice spacing really dominates the picture, we may at least have a look at the running couplings $\alpha_{\MSbar}(1/a)=g^2_{\MSbar}(1/a)/(4\pi)$ associated with the lattice spacings typically available, using perturbative 5-loop running~\cite{Luthe:2017ttc}.
A non-exhaustive overview of lattice spacings available in the literature is given in \tab{tab:typicalLatticeSpacings}, where one finds that at least the smallest lattice spacings clearly start to reach the perturbative region\footnote{We assume here that the upper bound of the perturbative region is roughly $\alpha_{\MSbar}(1/a)\sim 0.25$ beyond which a perturbative description will break down.}.
Of course, even smaller lattice spacings would (always) be preferable, but may be too costly at the moment.
Meanwhile step-scaling analysis in pure gauge theory~\cite{DallaBrida:2019wur} reach down to lattice spacings as small as $a<10^{-3}\,\mathrm{fm}$.
There, the absolute systematic effect on the extracted continuum value should become less severe when using the classical $a^n$ power law instead of including the logarithmic corrections discussed here.

Looking at table~\ref{tab:leadingPowers}, we can also get a glimpse of some contributions relevant to the use of the Relativistic Heavy Quark (RHQ) action~\cite{Christ:2006us} assuming $aM\ll 1$ for the heavy quark mass $M$.
There a heavy flavour, e.g.~the bottom quark, is implemented via a different lattice discretisation compared to the lighter quark flavours.
Although the RHQ action is based on the Wilson action, it breaks the spacetime symmetry even further by treating spatial directions differently than the Euclidean time direction, i.e.~we can no longer require hypercubic symmetry for our minimal basis.
Therefore, the leading powers in the coupling presented here give only an incomplete picture as different powers in the coupling will arise for the additional operators allowed by the less restrictive spacetime symmetries.

In case one is interested in non-spectral quantities each field introduces an additional minimal basis of higher dimensional operators compatible with the transformation behaviour of the local field.
The associated spectrum must then be taken into account as well.
An analysis of such additional powers in the coupling for fermion bilinears is on its way.
In general such computations aiming only at 1-loop anomalous dimensions should not be too complicated, but finding the minimal basis is somewhat tedious.
As mentioned before, once the minimal basis is found, the strategy described here should be applicable also for the local fields with the sole difference that the operators must be renormalised at non-zero momentum or in other words mixing with total divergence operators must be taken into account as well.
In particular for precision physics observables, performing these computations should become the norm as it offers better control over the continuum extrapolation by either reducing the error budget or providing a better estimate for this uncertainty.

A more complicated issue is the generalisation to staggered quarks~\cite{Kogut:1974ag}, which requires a whole new class of operators that allow for flavour changing interactions, see e.g.~\cite{Kilcup:1986dg}, and thus have a severely reduced flavour symmetry compared to Ginsparg-Wilson quarks, whose operators of course still contribute with the known values for $\hat{\Gamma}_i^{(2)}$.
For the additional set of operators we do not know whether they fall within the same range of values $\hat{\Gamma}_i^{(2)}$ or possibly undershoot them.
The latter should of course give rise to concerns as there is no theoretical lower bound.
Due to the prominence in the literature and low computational cost of this lattice discretisation in numerical simulations, this is probably the most pressing gap in this work.
However, this is not the only theoretical concern arising when using staggered quarks.
Firstly, there is no proof of perturbative renormalisability of staggered quarks to all orders in perturbation theory available, using the lattice power counting theorem generalised for staggered quarks~\cite{Giedt:2006ib}.
Such a proof exists for Wilson~\cite{Reisz:1988kk} and Ginsparg-Wilson fermions~\cite{Reisz:1999ck}.
Secondly, so called \textit{rooting} to reduce the number of flavours, makes rooted staggered quarks a highly non-local theory, see e.g.~\cite{Sharpe:2006re}.
Both aspects may invalidate the applicability of Symanzik Effective theory altogether.
Nonetheless a Symanzik Effective theory analysis can still be done keeping in mind that both issues mentioned here must be resolved independently to have a solid theoretical basis.

\appendix
\section{Conventions}

\label{app:latticederivatives}
The gauge covariant lattice forward and backward derivatives acting on quark fields are defined as
\begin{align}
\nabla_\mu \Psi(x)&=\frac{U(x,\mu)\Psi(x+a\hat{\mu})-\Psi(x)}{a},\\
\nabla_\mu^*\Psi(x)&=\frac{\Psi(x)-U^\dagger(x-a\hat{\mu},\mu)\Psi(x-a\hat{\mu})}{a},
\end{align}
where $U(x,\mu) \in $~SU($N$) are the link variables connecting $x+a\hat\mu$ and $x$.
For improvement of the Wilson Dirac operator we require a lattice discretisation of the field strength tensor, which we assume here to be the clover term
\begin{align}
a^2\hat{F}_{\mu\nu}&=\frac{1}{8}\left\{Q_{\mu\nu}(x)-Q_{\nu\mu}(x)\right\},\\
Q_{\mu\nu}(x)&=U(x,\mu)U(x+a\hat{\mu},\nu)U^\dagger(x+a\hat{\nu},\mu)U^\dagger(x,\nu)\nonumber\\
&+U(x,\nu)U^\dagger(x-a\hat{\mu}+a\hat{\nu},\mu)U^\dagger(x-a\hat{\mu},\nu)U(x-a\hat{\mu},\mu)\nonumber\\
&+U^\dagger(x-a\hat{\mu},\mu)U^\dagger(x-a\hat{\mu}-a\hat{\nu},\nu)U(x-a\hat{\mu}-a\hat{\nu},\mu)U(x-a\hat{\nu},\nu)\nonumber\\
&+U^\dagger(x-a\hat{\nu},\nu)U(x-a\hat{\nu},\mu)U(x+a\hat{\mu}-a\hat{\nu},\nu)U^\dagger(x,\mu).
\end{align}

\vskip 0.3cm

\noindent

\bibliographystyle{JHEP}

\providecommand{\href}[2]{#2}\begingroup\raggedright\endgroup

\end{document}